\documentclass[%
reprint,
 amsmath,amssymb,
 aps,
]{revtex4-2}

\usepackage{subfigure}
\usepackage{float}
\usepackage{slashed}
\usepackage{xcolor}
\usepackage{amsmath}
\usepackage{subcaption}
\usepackage[english]{babel}
\usepackage{lineno, blindtext}
\usepackage{tikz}
\usepackage{adjustbox}
\usepackage{rotating}
\usepackage{rotfloat}
\usepackage{verbatim}
\usepackage[utf8]{inputenc}
\usepackage{graphicx}
\usepackage{dcolumn}
\usepackage{bm}
\usepackage[super]{nth}
\usepackage{lipsum}

\newcommand\Tdiag[4]{%
    \multicolumn{1}{|p{#2}|}{\hskip-\tabcolsep
    \begin{tikzpicture}[%
                baseline={(0,-.25\baselineskip)},
                every node/.style={outer sep=0pt,inner sep=#1}]
    \node[minimum width={#2+1\tabcolsep-\pgflinewidth},
        minimum height=2\baselineskip-\pgflinewidth+\extrarowheight,
        use as bounding box] (box) {};
    \draw[line cap=round] (box.north west) -- (box.south east);
    \node[anchor=south west,text width=.75*#2,align=left] at (box.south west) {#3};
    \node[anchor=north east,text width=.75*#2,align=right] at (box.north east) {#4};
\end{tikzpicture}\hskip-\tabcolsep}}

\usepackage{hyperref}
\usepackage{xcolor}
\hypersetup{
  colorlinks   = true, 
  urlcolor     = red, 
  linkcolor    = blue, 
  citecolor   = blue 
}

\usepackage{threeparttable}
\begin{document}

\preprint{APS/123-QED}

\title{Angular distribution study for high mass dimuon pairs in CMS open 2012 data and for Mono-Z$^{\prime}$ model}

\author{S. Elgammal}
 \altaffiliation[sherif.elgammal@bue.edu.eg]{}
\affiliation{%
Centre for Theoretical Physics, The British University in Egypt, P.O. Box 43, El Sherouk City, Cairo 11837, Egypt.
}%


\date{\today}

\begin{abstract}
{This analysis uses the CMS open data to study the angular distribution of high-mass dimuon pairs produced during proton-proton interaction at a center of mass energy of 8 TeV. The study uses the cos$\theta_{\text{CS}}$ variable defined in the Collins Soper frame. In the Standard Model, the production of high-mass dimuon pairs is dominated by the Drell-Yan process, which exhibits a strong forward-backward asymmetry. 
However, many scenarios beyond the Standard Model predict different shapes for the cos$\theta_{CS}$ distribution. In excess events beyond the Standard Model, the angular distribution can be used to distinguish between those models.
The Mono-Z$^{\prime}$ model has been used to interpret the data, and no deviation from the SM has been observed.}

\vspace{0.75cm}
\end{abstract}

\maketitle






\section{Introduction}
\label{sec:intro}
One possible method of detecting physics beyond the Standard Model at the Large Hadron Collider is by observing changes in the dilepton mass spectrum at high mass. These changes could come in the form of a new peak, which is predicted by heavy neutral gauge boson models like Z$^{\prime}$ \cite{heaveyZ} or Randall-Sundrum \cite{extradim}, or a broad distortion of the spectrum, which could indicate the presence of Contact Interaction \cite{leptoquark, ContactInteraction} or ADD \cite{ADD} models. 

The CMS collaboration has closely studied these signatures, particularly to Z$^{\prime}$ and Contact Interaction models \cite{ZprimeandCI}. 
Another model, known as Mono-Z$^{\prime}$ \cite{R1}, predicts the production of dark matter in association with the Z$^{\prime}$. 
To confirm these theories, the mass spectrum must show an excess or a deficit of events compared to the background prediction, which is dominated by the Drell-Yan process. 
Additionally, the angular distributions of the leptons would be affected. 

Previous studies by the CMS \cite{AfbCMS} and ATLAS \cite{AfbATLAS} collaborations used the angular distributions of the Drell-Yan charged lepton pairs around the Z-boson mass peak to measure the forward-backward asymmetry $(A_{FB})$. Both studies have used the 
full LHC run 1 data, corresponding to an integrated luminosity of 19.7 $\text{fb}^{-1}$ for the CMS and 20.3 $\text{fb}^{-1}$ for the ATLAS, of pp collisions at 8 TeV center of mass energy $(\sqrt{s})$.
They found that the measurements of $A_{FB}$ are consistent with the standard model
predictions.
The forward-backward asymmetry of the high mass dilepton ($M_{ll} > 170$ GeV) has been measured in the CMS detector at $\sqrt{s}$ = 13 TeV with an integrated luminosity of
138 $\text{fb}^{-1}$, this analysis concluded that no statistically significant deviations from standard model predictions are observed \cite{{AfbCMS13tev}}.

In the current analysis, we examined the angular distributions of the dimuon channel
specifically focusing on mass bins higher than the Z-boson mass peak using the open data collected by the CMS experiment in 2012 during the LHC run 1 of data taking, corresponding to an integrated luminosity of 11.6 $\text{fb}^{-1}$ at $\sqrt{s} = 8$ TeV.
We used the Mono-Z$^{\prime}$ model to interpret the data, and statistical interpretation of the results is given.

The analysis is structured as follows:
Section \ref{section:CS} introduces the Collins-Soper frame, the variable cos$\theta_{\text{CS}}$, and describes its interpretation for Drell-Yan events. 
The ambiguity in the determination of the quark direction for such events is also discussed. 
In the following paper, we will discuss the theoretical model for the production of a neutral gauge boson (Z$^{\prime}$) in association with dark matter (DM) particles at the LHC. This will be covered in section \ref{section:model}. We will also briefly describe the CMS detector in section \ref{section:CMS}. 
In section \ref{section:MCandDat}, we will mention the CMS open data and Monte Carlo (MC) samples used in the analysis of proton-proton collisions. After that, we will cover important SM background processes and how to calculate their contributions in section \ref{section:Backgrounds}. Furthermore, the analysis strategy and criteria for the event selection will be discussed in section \ref{section:AnSelection}, followed by a section on systematic uncertainties and their effect on predicting the backgrounds in section \ref{section:Uncertainties}. 
In section \ref{section:Results}, the results of this analysis based on the angular distribution (cos$\theta_{\text{CS}}$) after the use of the event final selection and the statistical interpretation are presented.
Finally, we have shown the summary of the search in sections \ref{section:Summary}.
\section{The Collins-Soper frame}
\label{section:CS}
When two partons collide, they can produce a pair of leptons known as $l^+l^-$. To describe the distribution of these leptons, we use an angle called $\theta$, which measures the angle between the negative lepton and one of the partons in the dilepton pair's center of mass. In the Standard Model, the only process that occurs at the tree level to produce an $l^+l^-$ pair is the Drell-Yan process ($q\bar{q} \rightarrow \gamma^*/Z \rightarrow l^+l^-$).

Our analysis studies the angular distribution of lepton pairs produced when two hadrons collide, using the Collins-Soper (CS) frame \cite{CSpaper}. This frame minimizes the distortion of angular distributions caused by the transverse momenta of the two incoming partons. Measuring $\theta$ or cos$\theta$ exactly is impossible if at least one of the two incoming partons has a non-zero transverse momentum. Furthermore, the quark direction is not directly known in the case of a $q\bar{q}$ annihilation.

To investigate the angular distribution of lepton pairs, we utilize the Collins-Soper frame. This frame defines the angle $\theta_{CS}$ as the angle between the negative lepton momentum and the z-axis. Using this frame, we can minimize the uncertainties caused by the unknown transverse momentum of the incoming quarks.

To determine the orientation of the Collins-Soper frame, we rely on the sign of the longitudinal boost of the dilepton system. We can compute the angle $\text{cos} \theta_{CS}$ from quantities that we measure in the lab frame, as explained in \cite{AfbCMS}.
\begin{equation}
     \text{cos}\theta_{CS} = \frac{|Q_z|}{Q_z} \frac{2(P_1^+ P_2^- - P_1^- P_2^+)}{\sqrt{Q^2(Q^2 + Q_T^2)}}.     
    \label{costheta:equ}
\end{equation}
The symbols $Q$, $Q_T$, and $Q_z$ stand for the four-momentum, the transverse momentum, and the longitudinal momentum of the dimuon system, respectively. Similarly, $P_1$ ($P_2$) represents the four-momentum of $\mu^-$ ($\mu^+$), and $E_i$ denotes the energy of the muon. In addition, $P^\pm_i$ is defined as $(E_i \pm P_{z,i})/\sqrt{2}$.

\begin{figure} [h!]
\centering
\resizebox*{8.0cm}{!}{\includegraphics{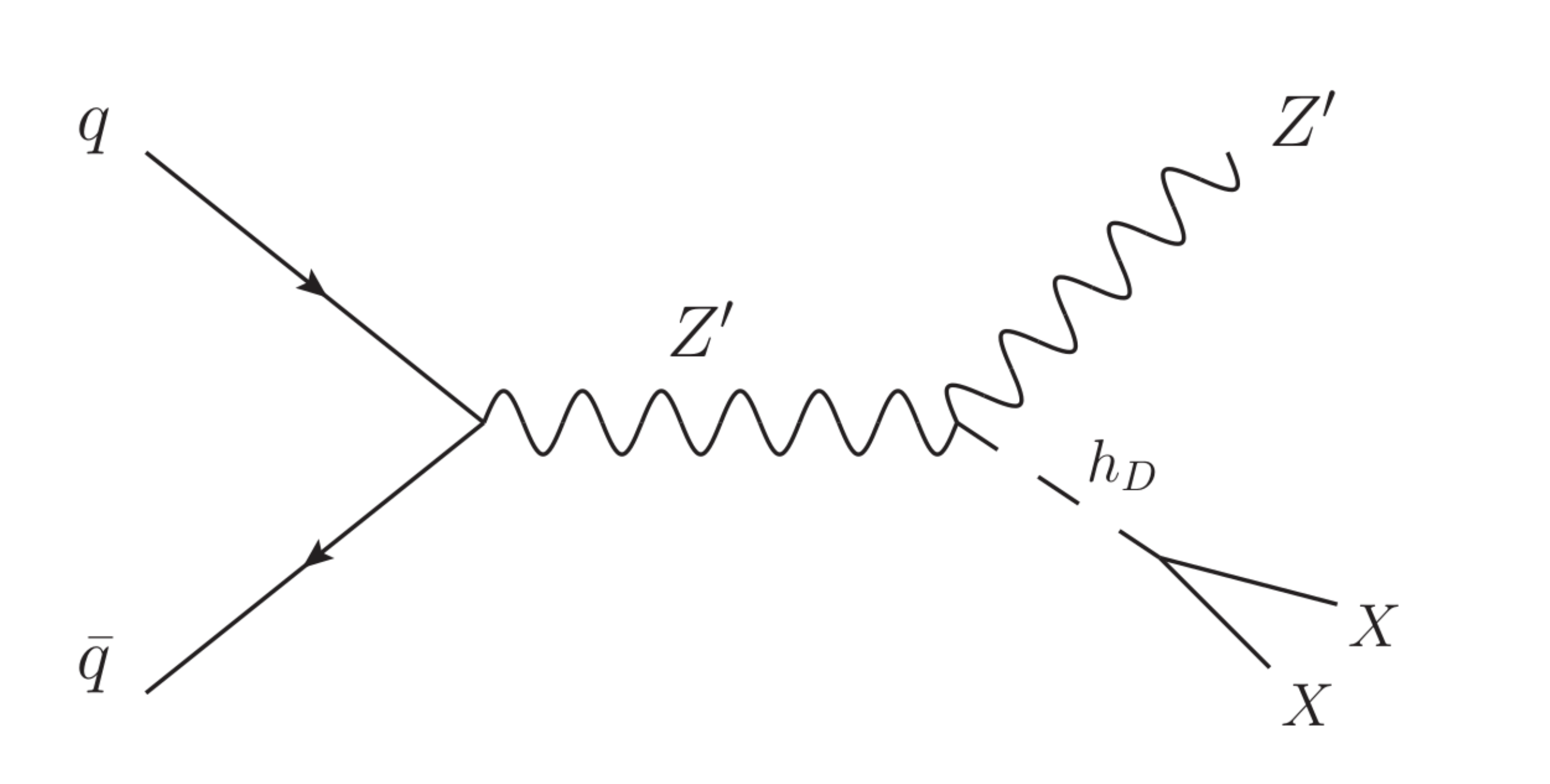}}
\caption{The Feynman diagram of the dark Higgs scenario.}
\label{fig1}
\end{figure}
\section{The Mono-Z$^{\prime}$ model}
\label{section:model}
The model proposed in reference \cite{R1} suggests the generation of dark matter particles and a neutral gauge boson called Z$^{\prime}$ through a $q\bar{q}$ annihilation process in pp collisions. The mediator vector boson, Z$^{\prime}$, interacts with the Standard Model particles and dark matter particles in the simplified Dark Higgs scenario. The Z$^{\prime}$ boson generates a dark sector Higgs ($h_{D}$), which then disintegrates into a pair of DM particles ($\chi \bar{\chi}$) assuming the masses of the dark Higgs and Z$^{\prime}$ are equal. The mass choice is given in table \ref{table:tab1} and the Feynman diagram is shown in figure \ref{fig1}.

\begin{table} [h!]    
\centering
\begin {tabular} {ll}
\hline
\hline
Scenario & \hspace{0.8cm} Masses assumptions \\
\hline
\\
    Heavy dark sector & \hspace{0.8cm}  $M_{h_{D}} =$ 
    $\begin{cases}
        125~\text{GeV}, & M_{Z'} < 125~\text{GeV} \\
        M_{Z'}, & M_{h_{D}} > 125~\text{GeV}.
  \end{cases}$\\
 & \\
\hline
\hline
\end {tabular}
\caption{The assumptions of the masses of the particles produced following the DH scenario, the heavy dark sector introduced in \cite{R1}.}
\label{table:tab1}
\end{table}
\begin{table*}
\centering
\fontsize{8.pt}{12pt}
\selectfont
\begin{tabular}{|c|c|c|c|c|c|c|c|c|c|}
\hline
\Tdiag{.4em}{2.1cm}{$M_{\chi}$}{$M_{Z'}$}  & 200 & 250 & 300 & 350 & 400 & 450 & 500 & 600 &700  \\
\hline
5  & 
$2.37\times10^{-02}$&  
$0.96\times10^{-02}$& 
$0.44\times10^{-02}$&  
$0.21\times10^{-02}$&  
$0.10\times10^{-02}$&  
$0.57\times10^{-03}$&  
$0.33\times10^{-03}$&  
$1.19\times10^{-04}$&  
$4.73\times10^{-05}$
\\
\hline
\end {tabular}
\caption{Cross-sections times branching ratios in pb for the DH scenario, calculated for different values of $M_{Z'}$ in (GeV), assuming $M_{\chi}$ = 5 GeV, coupling constant $g_{DM}$ = 1.0, and $g_{SM}$ = 0.25, at $\sqrt{s} = 8$ TeV \cite{SherifMagdy}.}
\label{table:tabchi}
\end{table*}
We have conducted a study on the DH simplified model, which involves free parameters such as mediator mass ($M_{Z'}$), masses of dark Higgs ($M_{h_{D}}$), and couplings between the mediator and SM ($g_{SM}$) and DM fields ($g_{DM}$). Our research focuses on the Mono-Z$^{\prime}$ process, which produces a dimuon in addition to missing transverse momentum, belonging to DM candidates. We have simulated Mono-Z$^{\prime}$ samples for mediator masses ranging from 150 GeV to 700 GeV, with DM mass ($M_{\chi} = 5$ GeV) and coupling constants $g_{SM} = 0.25$ and $g_{DM} = 1.0$. The values of these coupling constants are recommended by \cite{R37} to obtain the highest possible cross-sections multiplied by the branching ratios. 
In a study by Ref. \cite{R37}, it was determined that the coupling constant values $g_{SM}$ are excluded in the range of 0.02–0.2 for Z$^{\prime}$ masses between 200 and 1000 GeV in heavy dark-sector scenarios in the context of the dark-Higgs model.

The cross-sections multiplied by the branching ratios are calculated using MadGraph5 aMC@NLO v2.6.7 \cite{R33} at next-to-leading order, regarding pp collisions at the LHC with 8 TeV center of mass energy, for the DH scenario. They are listed in table \ref{table:tabchi}. 
It was found in \cite{SherifMagdy} that cross-section times the branching ratio varies with the change of both the mediator and dark Higgs masses and does not depend on the choice of dark matter mass. 
We have also calculated the decay widths \cite{dwidth, R33} for each Z$^{\prime}$ and the $h_{D}$ for the DH scenario \cite{SherifMagdy}.
\section{The CMS detector}
\label{section:CMS}
The CMS detector \cite{R29} is a particle detector at the LHC, designed to investigate a range of physics including the Higgs boson, dark matter, and extra dimensions. It has five layers, each with sub-detectors and a superconducting solenoid: inner tracker, ECAL, HCAL, magnet, and muon system.
The CMS coordinate system has the z-axis aligned with the beam axis, the y-axis pointing upwards, and the x-axis pointing towards the center of the LHC. The azimuthal angle ($\phi$) is the angle in the transverse plane, while the polar angle ($\theta$) is measured from the positive z-axis and expressed in terms of pseudo-rapidity ($\eta$), where $ \eta = - \text{ln}[\text{tan}(\theta/2)] $.

In this study, we analyze muons and missing transverse energy. Muon objects are reconstructed by fitting muon tracks from both the inner tracker and the muon system, resulting in the production of global muons \cite{R18, R40, R19}.

The missing transverse momentum is calculated using the particle flow algorithm, which measures the imbalance in the vector sum of momenta in the transverse plane. It is the sum of negative PF reconstructed transverse momenta of all particles, as shown in the equation $\vec{\slashed{p}}_{T} = - \sum \vec{p}_{T}^{~\text{pf}}$. 
The value of $\vec{\slashed{p}}_{T}$ can be affected by various factors, which can be corrected using jet energy corrections, as defined in the formula provided in \cite{R45}.
$$ 
\vec{\slashed{p}}_{T}^{~\text{corr}} = \vec{\slashed{p}}_{T} -
\sum_{jets} (\vec{p}_{T jet}^{~\text{corr}} - \vec{p}_{T jet}).
$$

We are analyzing two variables: $\vec{\slashed{p}}_{T}^{~\text{corr}}$ and its magnitude $\slashed{\slashed{p}}_{T}^{~\text{corr}}$. The corrected values are labeled with "corr" suffix, and the latter is included in the Particle Flow (PF) MET object in the CMS software. References for the PF MET object are \cite{R45new} and \cite{pfmetopendata}.

\section{Data and simulated samples}
\label{section:MCandDat}
\subsection{Monte Carlo simulation of the model signals}
We generated the events of the signal model using $\text{MadGraph5\_aMC@NLO v2.6.7}$. 
The cross-section is calculated at NLO, and Pythia \cite{R34} is used for the hadronization process. The NNPDF2.3QED NLO set is used for the PDF \cite{PDFsignal}, 
and the detector simulation is done using the CMS open data software framework (the release $CMSSW\_5\_3\_32$ \cite{CMSSWversion}) at $\sqrt{s} = 8$ TeV. The effect of pileup has been simulated by overlaying MC-generated minimum bias events \cite{pileup}.
We scanned the production cross-section at different sets of masses for the Z$^{\prime}$ boson and dark matter particle $M_{\chi} = 5$ GeV, assuming $g_{SM}$ = 0.25 and $g_{DM}$ = 1.0.
\subsection{Monte Carlo simulation of the SM backgrounds}
To account for interference from SM processes with muons and/or missing energy, we used CMS open Monte Carlo samples at $\sqrt{s}$ = 8 TeV. The dominant background, Drell-Yan (DY), was generated using POWHEGBox v1.0 \cite{powheg1}, interfaced with the Pythia v.6.4.26 parton shower model. 
While fully leptonic decay of $\text{t}\bar{\text{t}}$ and diboson channels were generated using MadGraph5\_aMC@NLO interfaced with Pythia v.6.4.26. 
Additionally, the productions of electroweak diboson channels, such as WW and WZ, were generated using MadGraph interfaced with Pythia v.6.4.26. The ZZ to four muons process was also generated with POWHEGBox v1.0.
The Monte Carlo samples used in this analysis and their corresponding cross sections are listed in table \ref{table:tab3}.
The single-top background is missed in this analysis since it is not included in the list of the CMS open MC samples.

\begin{table*} 
\centering
\begin {tabular} {|l|l|l|c|l|}
\hline
Process \hspace{1cm} & Generator  & Data Set Name & {$\sigma \times \text{BR} ~(\text{pb})$} & Order \\
\hline
\hline
$\text{DY} (\mu\bar{\mu})$  & POWHEG &DYToMuMu\_M-20\_CT10\_TuneZ2star\_v2\_8TeV. \cite{R22}  & 1916 \cite{R3}& NNLO\hspace{6cm}\\
\hline
$\text{t}\bar{\text{t}}$ + jets & MADGRAPH  & TTJets\_FullLeptMGDecays\_8TeV. \cite{R23} & 23.89 \cite{ttbar}& NLO \\
\hline
WW + jets & MADGRAPH & WWJetsTo2L2Nu\_TuneZ2star\_8TeV. \cite{R24} & 5.8 \cite{R3}& NLO \\
\hline
WZ + jets & MADGRAPH  & WZJetsTo3LNu\_8TeV\_TuneZ2Star. \cite{R25} &1.1 \cite{R3}& NNLO \\
\hline
$ZZ\rightarrow 4\mu$  & POWHEG & ZZTo4mu\_8TeV. \cite{R26} & 0.077 \cite{R3}& NLO \\
\hline
\end {tabular}
\vspace{3pt}
\caption{The CMS used open MC samples to simulate the background events in proton-proton collisions at 8 TeV center-of-mass energy. The generators used and the MC samples' names were provided. The calculated cross sections $\times$ branching ratios, and order of calculations for each process are presented \cite{SherifMagdy}.}
\label{table:tab3}
\end{table*}
\subsection{CMS open data samples}
\label{section:CMSopenData}
The analysis is based on data files that stem from pp collisions at a center-of-mass energy of 8 TeV during the LHC run-I. These data files were recorded by the CMS detector in 2012. For this analysis, we used two open data runs, run-B, and run-C, with a total integrated luminosity of 11.6 fb$^{-1}$ \cite{Ropendata}. The high-level trigger HLT$\_$Mu40$\_$eta2p1 \cite{HLT} was employed to collect data. This trigger was un-prescaled for the full 2012 dataset and aimed to collect events with at least one muon candidate within $|\eta| <$ 2.1 and $\text{p}_{\text{T}} > 40$ GeV. The efficiency of this trigger varies as a function of $\eta$, resulting in efficiency for triggering on a dimuon system that ranges between 97\% and 100\% \cite{zprime}. The listed events are from the "good runs list" for the primary 2012 data set provided by the open data project \cite{R39}. 
Table \ref{table:tab4} presents the samples, run numbers, data set names, and corresponding integrated luminosity ($\mathcal{L}$).

 \begin{table*} 
\centering
\label{ tab-marks }
\begin {tabular} {|c|l|c|}
\hline
Run & Data Set & $\mathcal{L}$ ({fb}$^{-1}$) \\
\hline
\hline
Era B~ & SingleMu/Run2012B-22Jan2013-v1/AOD.\cite{R27}  &  \\ 
   &   & 11.6 \cite{Ropendata}\\
Era C~ & SingleMu/Run2012C-22Jan2013-v1/AOD.\cite{R28}  & \\
\hline
\end {tabular}
\vspace{2pt}
\caption{The used CMS open data samples during 2012 and the corresponding integrated luminosity \cite{SherifMagdy}.}
\label {table:tab4}
\end{table*}
\section{Backgrounds estimation}
\label{section:Backgrounds}
Three main types of SM backgrounds are considered in the search for new physics in the dimuon channel. The most significant background is from the fundamental SM Drell-Yan process. The second most important background comes from muons produced by non-singularly produced W and Z bosons, with $\text{t}\bar{\text{t}}$ events being the dominant source. The backgrounds from DY and other sources such as the leptonic decay of $t\bar{t}$, WW, WZ, and ZZ are estimated directly from Monte Carlo simulation and normalized to their corresponding cross-sections.

The third background source is known as the multijet background. 
It is caused by jets being misidentified as muons, which usually happens due to W+jets and QCD. 
Previous studies have estimated this background using a data-driven method in \cite{zprime}. They found that the jet misidentification resulted in 150 events representing approximately 0.15\% of the total SM backgrounds (98,596 events) 
estimated in the mass bin above 120 GeV \cite{zprime}. 
Which can have a very tiny effect on our results. For this reason, we neglect QCD and W+jets backgrounds estimated from the data.


In addition, we can suppress the cosmic muon contribution by constraining the vertex position and controlling the impact parameter of muons relative to the vertex position \cite{zprime}. 

\section{Event selection and systematic uncertainties}
\label{section:AnSelection}
\subsection{Selection of event}
\label{section:Preliminary}
We have established selection criteria for identifying muon candidates, which include: high transverse momentum ( $p^{\mu}_{T} > 45$ GeV), a limited range of $|\eta^{\mu}| < 2.1$, and global muons that pass the tracker-only isolation cut. To eliminate cosmic muons, we apply two cuts on the muon's transverse impact parameter and the 3D angle between the muon pairs. Additionally, we apply a $\chi^{2}/\text{dof}$ cut with a limit of 10 to ensure accurate pairing of muons and reject pile-up muons. 
These cuts are recommended by \cite{zprime}, and summarized in table \ref{table:selection1}. 
Finally, to decrease contamination of QCD and W+jets backgrounds, we select events with two opposite charge high $p_{T}$ muons passing the single muon trigger HLT$\_$Mu40$\_$eta2p1. We also apply cuts related to the dimuon invariant mass and $|\text{cos}\theta_{\text{CS}}|$, which must be less than 0.8 since we are not interested in events at which one of the two muons is very close to the beam pipe.

\begin{table*}
    \centering
    \begin {tabular} {|c|c|c|}
\hline
step & variable & requirements \\
\hline
    \hline
    & Trigger     & HLT\_Mu40\_eta2p1 \cite{HLT} \\
    & High $p_{T}$ muon ID & \cite{R41, R32}\\
Selection (1)    & $p^{\mu}_{T}$ (GeV) & $>$ 45 \\
    & $\eta^{\mu}$ (rad) & $<$ 2.1 \\
    &Mass window (GeV) & $(0.9 \times M_{Z^{\prime}}) < M_{\mu^{+}\mu^{-}} < (M_{Z^{\prime}} + 25)$ \\
    & $|\text{cos}\theta_{\text{CS}}|$& $\leq 0.8$ \\
\hline
    \end{tabular}
    \caption{Summary of selection (1) for the analysis.}
    \label{table:selection1}
\end{table*}
\subsection{Systematic uncertainties}
\label{section:Uncertainties}
The analysis takes into account different sources of uncertainties, both systematic and theoretical. Systematic uncertainties are related to experimental factors such as luminosity, muon detector acceptance, muon reconstruction efficiency, transverse momentum resolution, and detector geometry misalignment. On the other hand, theoretical uncertainties arise from the choice of parton distribution functions (PDF). The Drell-Yan cross-section calculation's uncertainty varies depending on the dimuon invariant mass. For the ZZ and WZ processes, the PDF uncertainties were 5\% and 6\%, respectively. 
To summarize, table \ref{table:sources} provides an overview of the sources of uncertainty and their corresponding values.

\begin{table}[h!]
    \centering
    \begin{tabular}{l r}
    \hline
    \hline
    Source     & Uncertainty (\%) \\
    \hline
    Luminosity ($\mathcal{L}$) & 2.6 {\footnotesize \cite{Lumi}} \\
    
    $A\times\epsilon$     & 3 {\footnotesize \cite{zprime}}\\
    
    $P_{T}$ resolution & 5 {\footnotesize \cite{zprime}}\\
    
    $P_{T}$ scale & 5 {\footnotesize \cite{zprime}}\\
   
    
    
    
    PDF (Drell-Yan) & 4.5 {\footnotesize \cite{zprime}}\\
    
    PDF (ZZ) & 5 {\footnotesize \cite{R450}}\\
    
    PDF (WZ)  & 6 {\footnotesize \cite{R450}}\\ 
    \hline
    \hline
    \end{tabular}
    \caption{Different sources of systematic uncertainties estimated and the corresponding values \cite{SherifMagdy}.}
    \label{table:sources}
\end{table}

\begin{figure}[h]
\centering
\resizebox*{10cm}{!}{\includegraphics{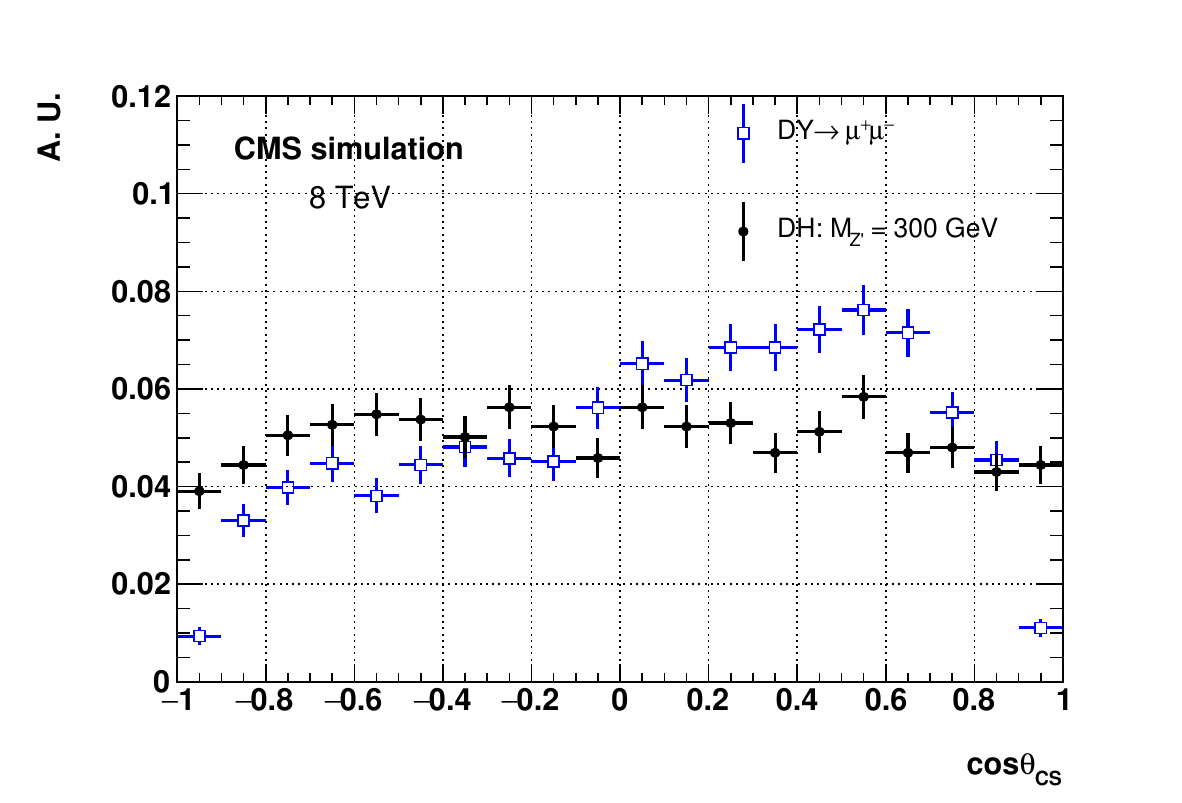}}
\caption{Normalized distributions of cos$\theta_{CS}$ for one resonant model mono-Z$^{\prime}$ (DH scenario) generated with a mass of 300 GeV and Drell-Yan events generated with a mass above 20 GeV. 
All events are required to pass the cuts listed in table \ref{table:selection1} (except cos$\theta_{CS}$ cut) and to have a reconstructed invariant mass in the range 270 - 325 GeV.
All histograms are normalized to unity to highlight qualitative features.} 
\label{fig2}
\end{figure}

We present the distribution of cos$\theta_{CS}$ for a resonant model, namely mono-Z$^{\prime}$ (DH scenario), in figure \ref{fig2}. The model has a generated mass of 300 GeV, and we compare it with Drell-Yan events. All events are required to satisfy the cuts mentioned in table \ref{table:selection1} (except cos$\theta_{CS}$ cut) and have a reconstructed invariant mass between 270-325 GeV. The results are depicted using black solid circles for the DH scenario and blue open boxes for Drell-Yan events, both normalized to unity. We observe a clear distinction between the DH model and the Drell-Yan events.
\begin{figure*}[]
\centering
\subfigure[135 $< M_{\mu^+\mu^-} <$ 175 GeV]{
  \includegraphics[width=80mm]{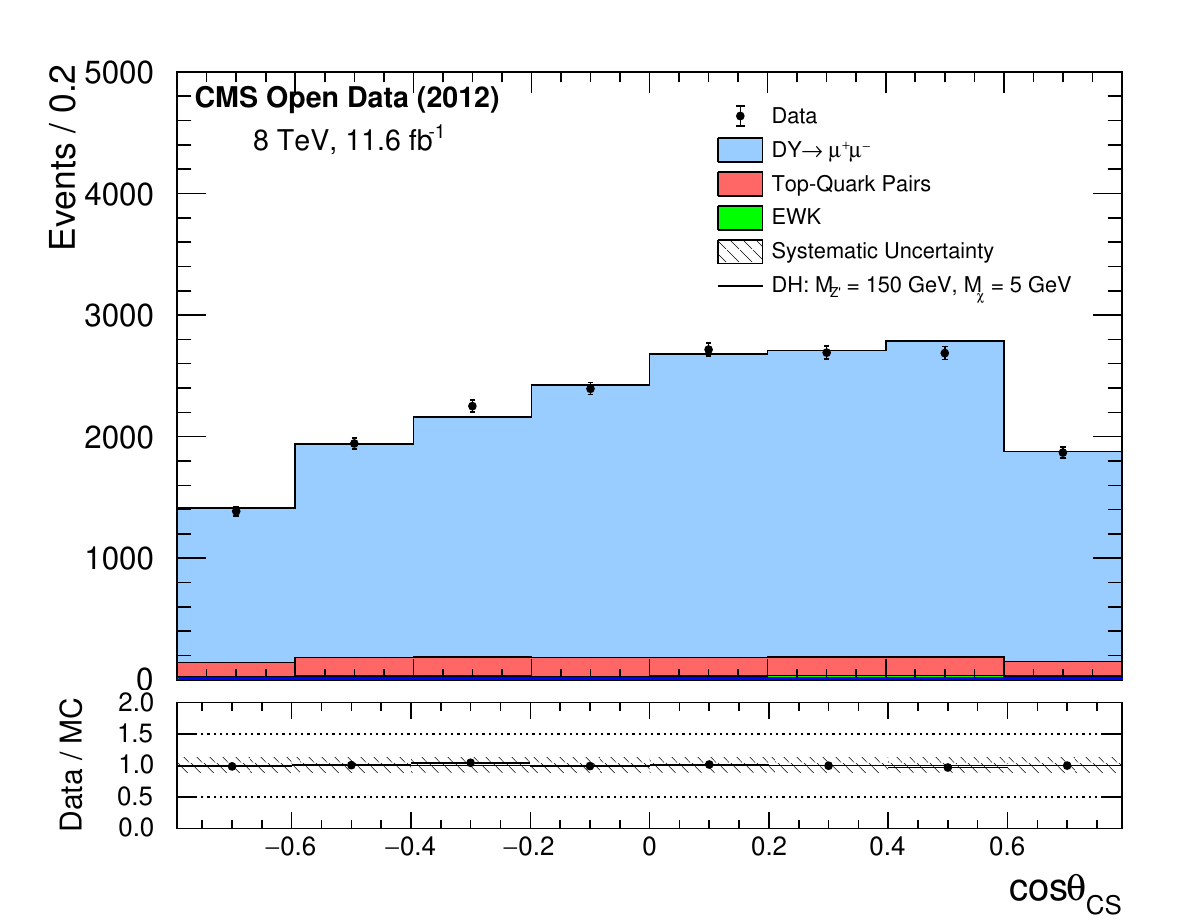}
  \label{bin1}
}
\hspace{0mm}
\subfigure[180 $< M_{\mu^+\mu^-} <$ 225 GeV]{
  \includegraphics[width=80mm]{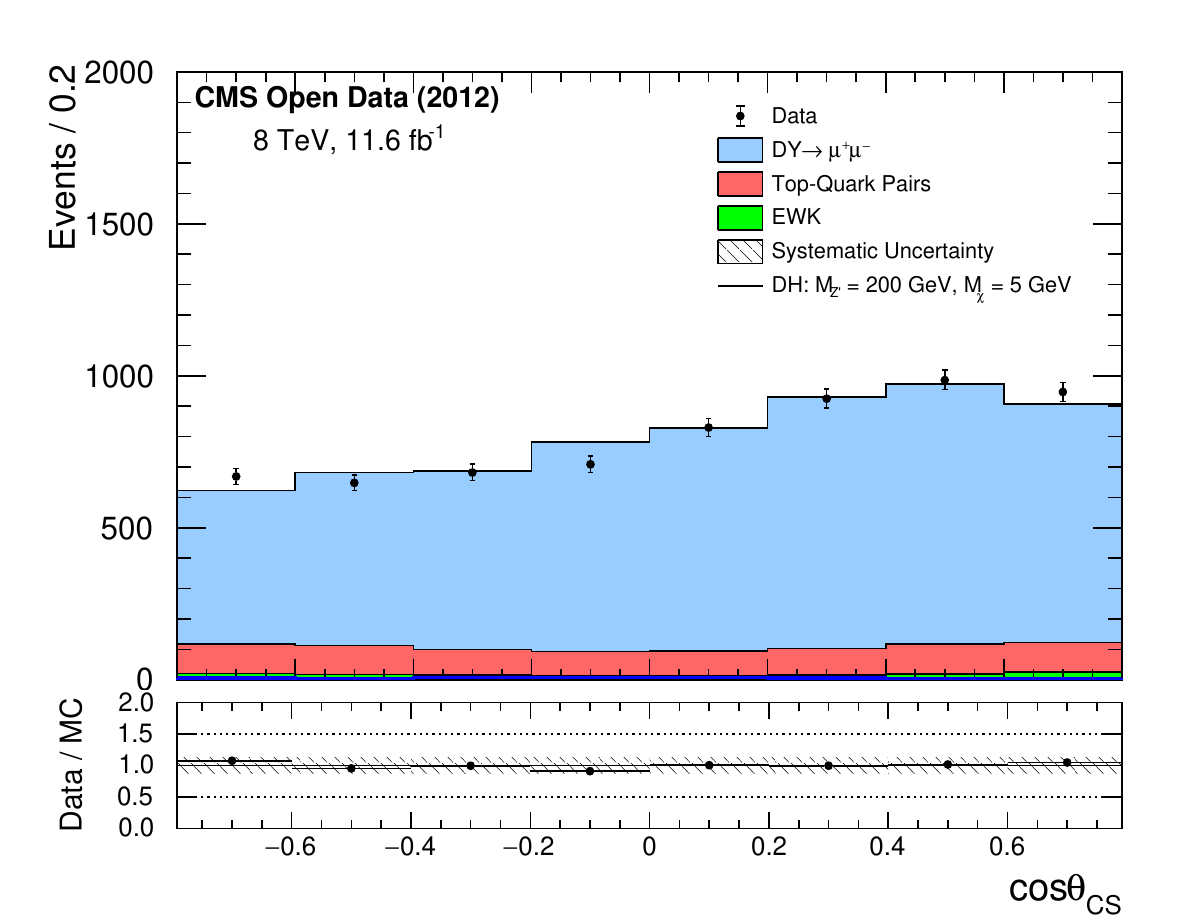}
  \label{bin2}
}
\hspace{0mm}
\subfigure[270 $< M_{\mu^+\mu^-} <$ 325 GeV]{
  \includegraphics[width=80mm]{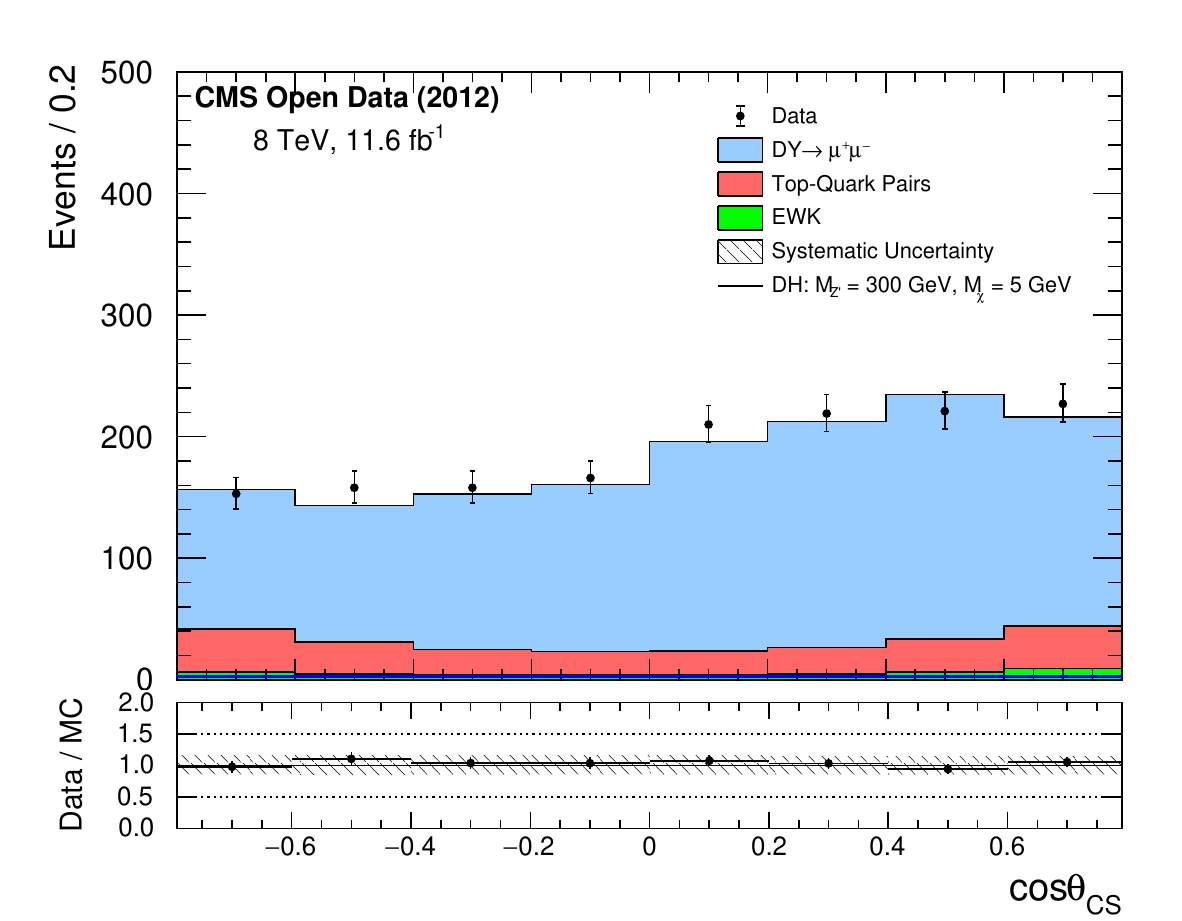}
  \label{bin3}
}
\hspace{0mm}
\subfigure[360 $< M_{\mu^+\mu^-} <$ 425 GeV]{
  \includegraphics[width=80mm]{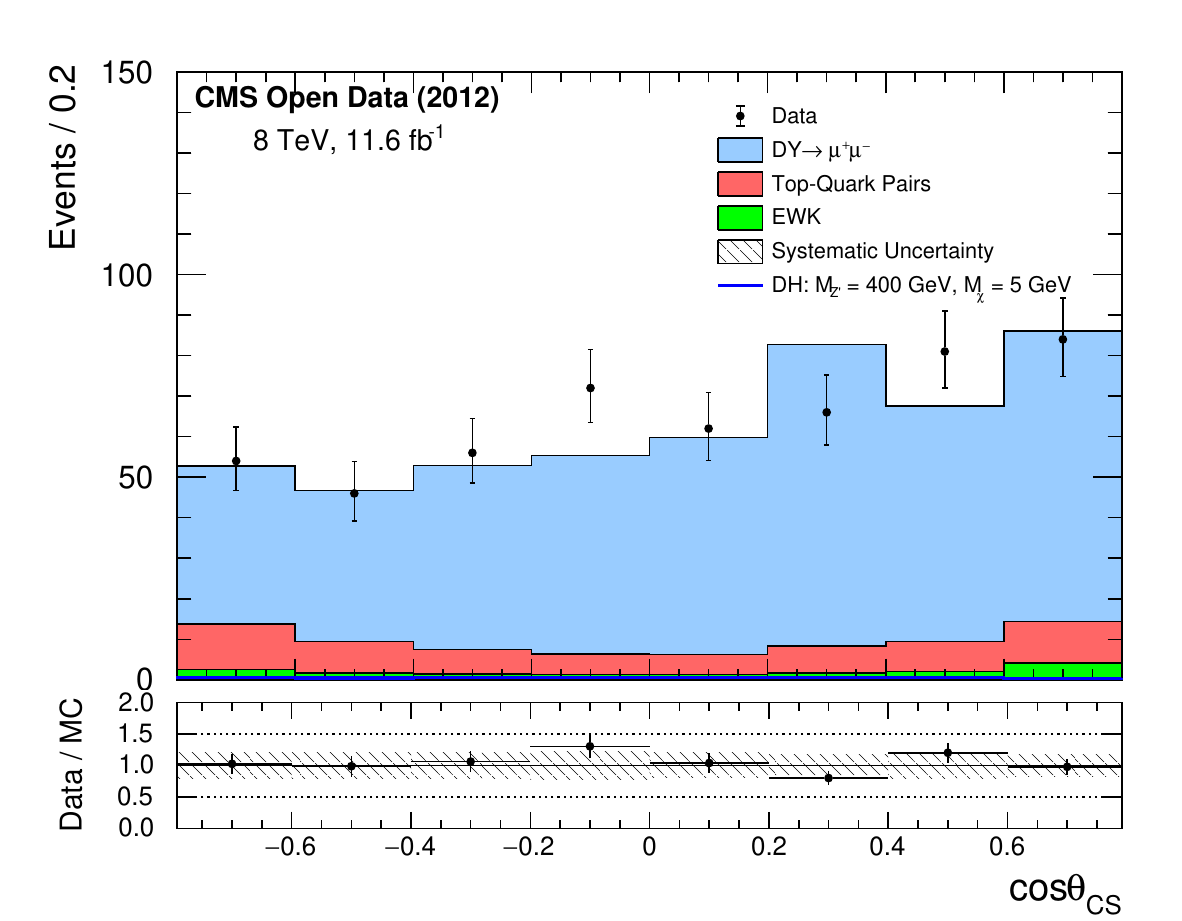}
  \label{bin4}
}
\hspace{0mm}
\subfigure[450 $< M_{\mu^+\mu^-} <$ 525 GeV]{
  \includegraphics[width=80mm]{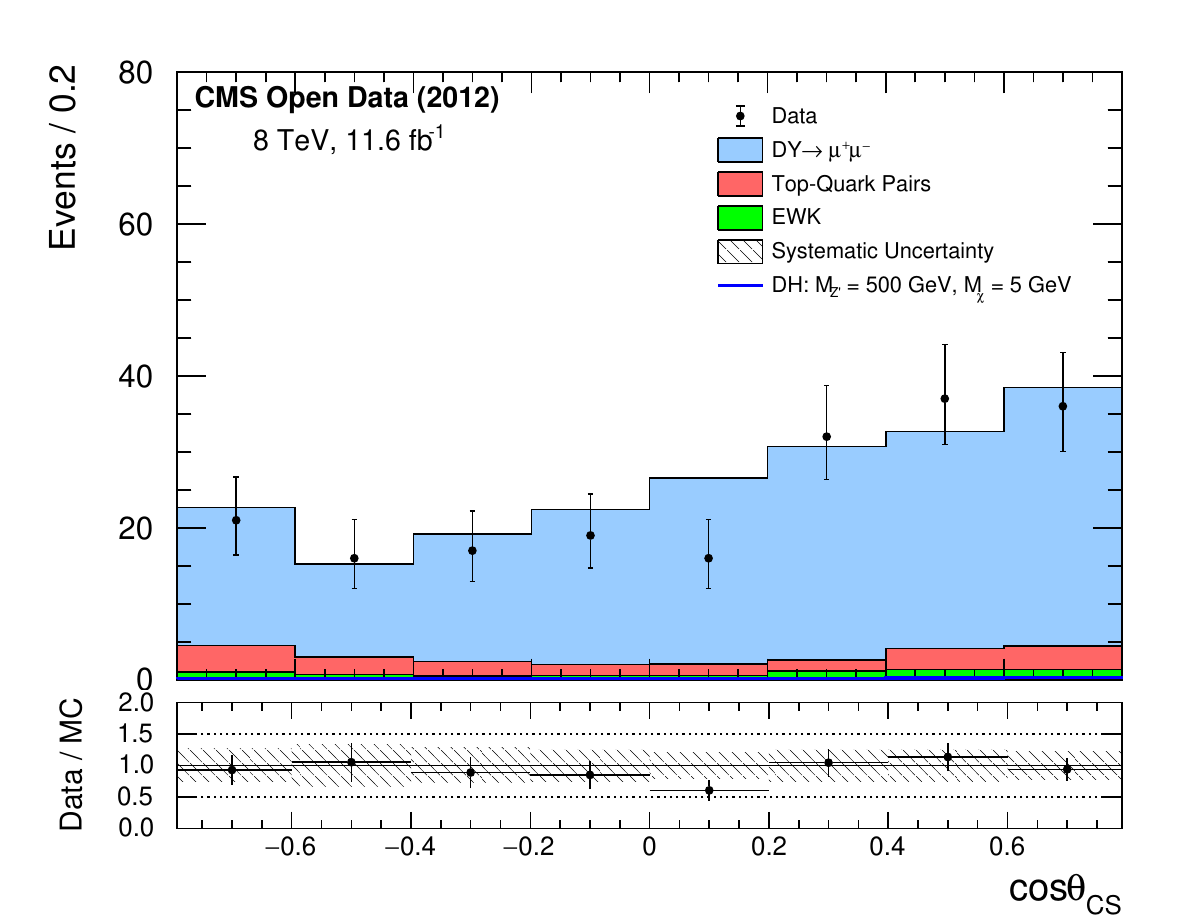}
  \label{bin5}
}
\hspace{0mm}
\subfigure[540 $< M_{\mu^+\mu^-} <$ 625 GeV]{
  \includegraphics[width=80mm]{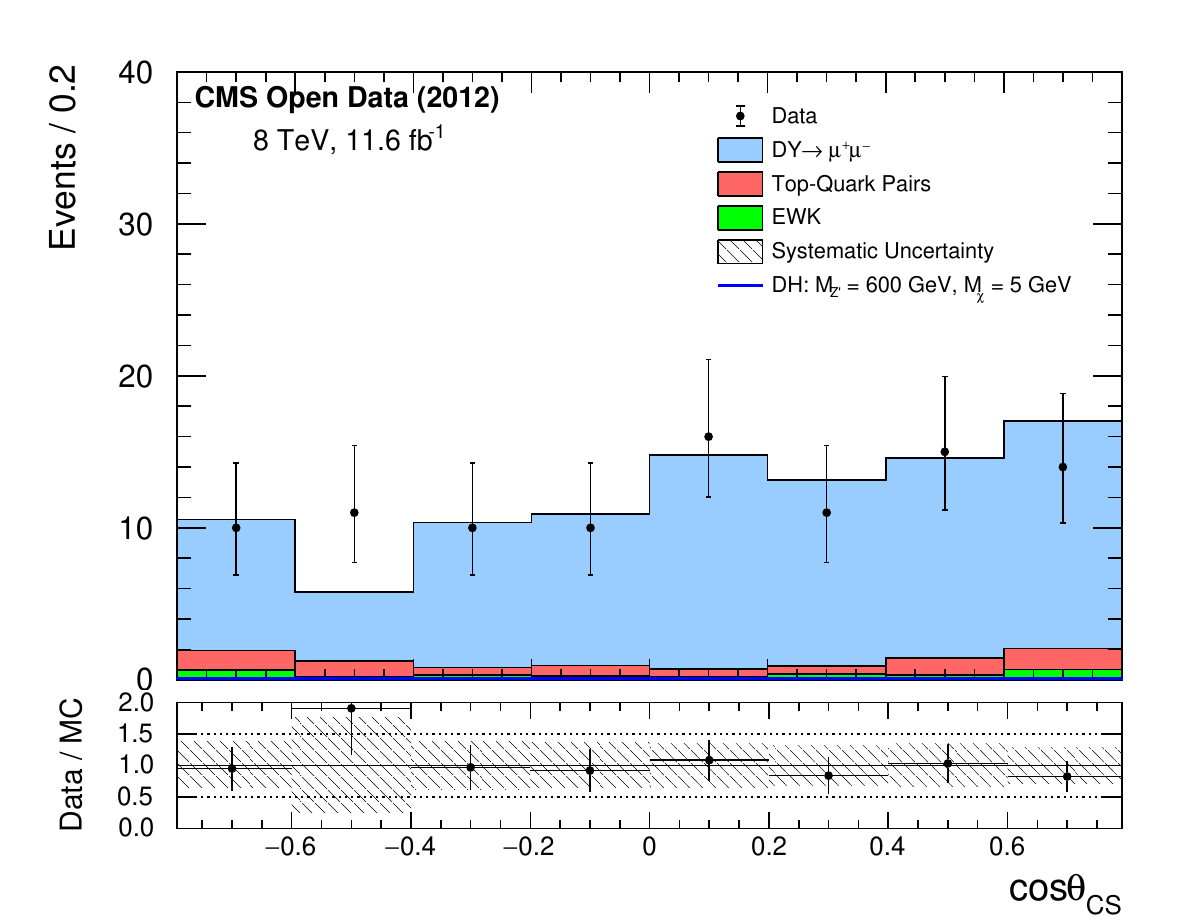}
  \label{bin6}
}
\caption{Distributions of cos$\theta_{CS}$ are illustrated, for events passing selection 1 listed in table \ref{table:selection1}, for data (dots) and Standard Model expectations (histograms) for several dimuon mass windows: 
135 $< M_{\mu^+\mu^-} <$ 175 GeV (\ref{bin1}), 
180 $< M_{\mu^+\mu^-} <$ 225 GeV (\ref{bin2}), 
270 $< M_{\mu^+\mu^-} <$ 325 GeV (\ref{bin3}), 
360 $< M_{\mu^+\mu^-} <$ 425 GeV (\ref{bin4}), 
450 $< M_{\mu^+\mu^-} <$ 525 GeV (\ref{bin5}) and 
540 $< M_{\mu^+\mu^-} <$ 625 GeV (\ref{bin6}). 
The signals presentation of the model corresponding to the DH scenario with the value of $M_{Z^{\prime}}$ runs from 150 to 600 GeV are superimposed.
In the lower bands, the ratio between the data and simulation is shown and the shaded region corresponds to the statistical and systematic uncertainties in the predicted backgrounds, added in quadrature.}
\label{fig3}
\end{figure*}

The graphs presented in figure \ref{fig3} illustrate the distribution of the cos$\theta_{CS}$ for various mass bins 
(135 $< M_{\mu^+\mu^-} <$ 175 GeV (\ref{bin1}), 
180 $< M_{\mu^+\mu^-} <$ 225 GeV (\ref{bin2}), 
270 $< M_{\mu^+\mu^-} <$ 325 GeV (\ref{bin3}), 
360 $< M_{\mu^+\mu^-} <$ 425 GeV (\ref{bin4}), 
450 $< M_{\mu^+\mu^-} <$ 525 GeV (\ref{bin5}), and 
540 $< M_{\mu^+\mu^-} <$ 625 GeV (\ref{bin6}))
of events that have passed the selection criteria mentioned in table \ref{table:selection1}. The black solid dots with error bars represent the CMS open data for an integrated luminosity of 11.6 fb$^{-1}$, showing only the statistical error. The cyan, red, and green histograms represent the Drell-Yan background, $t\bar{t}$ + jets background, and vector boson pair backgrounds (WW, WZ, and ZZ), respectively. These histograms are stacked, and the blue line displays the signal of the dark Higgs scenario generated with various masses of the neutral gauge boson Z$^{\prime}$ and $M_{\chi} = 5$ GeV. The lower bands show the data-to-simulation ratio. The shaded region indicates the statistical and systematic uncertainties (summarized in table \ref{table:sources}) in the predicted backgrounds, added in quadrature.

The data points and the simulated SM backgrounds in all the mass bins have shown good agreement in these plots. This is evident from the error bars on the data points, which demonstrate the statistical error, and the hatched region in the ratio plots, which represent the systematic uncertainty. 
However, since the signal samples are overwhelmed by the backgrounds, applying a tighter set of cuts is necessary to distinguish signals from SM backgrounds. 
The details of these cuts will be explained in section \ref{section:Results}.

\section{Results}
\label{section:Results}
After applying the first set of cuts presented in table \ref{table:selection1} the DH signal events were fully impeded in the SM background. In addition, a tighter selection has been optimized to distinguish DH signals from the SM backgrounds. 
The tight selection is based on three variables: 
A strong cut on the missing transverse momentum ($\slashed{p}_{T}^{\text{corr}} > 150$ GeV).
The azimuthal angle difference between the dimuon system and the missing transverse momentum $\Delta\phi_{\mu^{+}\mu^{-},\vec{\slashed{p}}_{T}^{\text{corr}}}$.
The relative difference between the dimuon system transverse momentum and the missing transverse momentum $|p_{T}^{\mu^{+}\mu^{-}} - \slashed{p}_{T}^{\text{corr}}|/p_{T}^{\mu^{+}\mu^{-}}$.
Here, $p_{T}^{\mu^{+}\mu^{-}}$ is the dimuon transverse momentum, and $\Delta\phi_{\mu^{+}\mu^{-},\vec{\slashed{p}}_{T}^{\text{corr}}}$ is defined as the difference in the azimuth angle between the dimuon system direction and the missing transverse momentum direction (i.e., $\Delta\phi_{\mu^{+}\mu^{-},\vec{\slashed{p}}_{T}^{\text{corr}}} = |\phi^{\mu^{+}\mu^{-}}-~\phi^{miss}|$ ).
\begin{figure*}
\centering
\subfigure[$\slashed{p}_{T}^{\text{corr}}$]{
  \includegraphics[width=85mm]{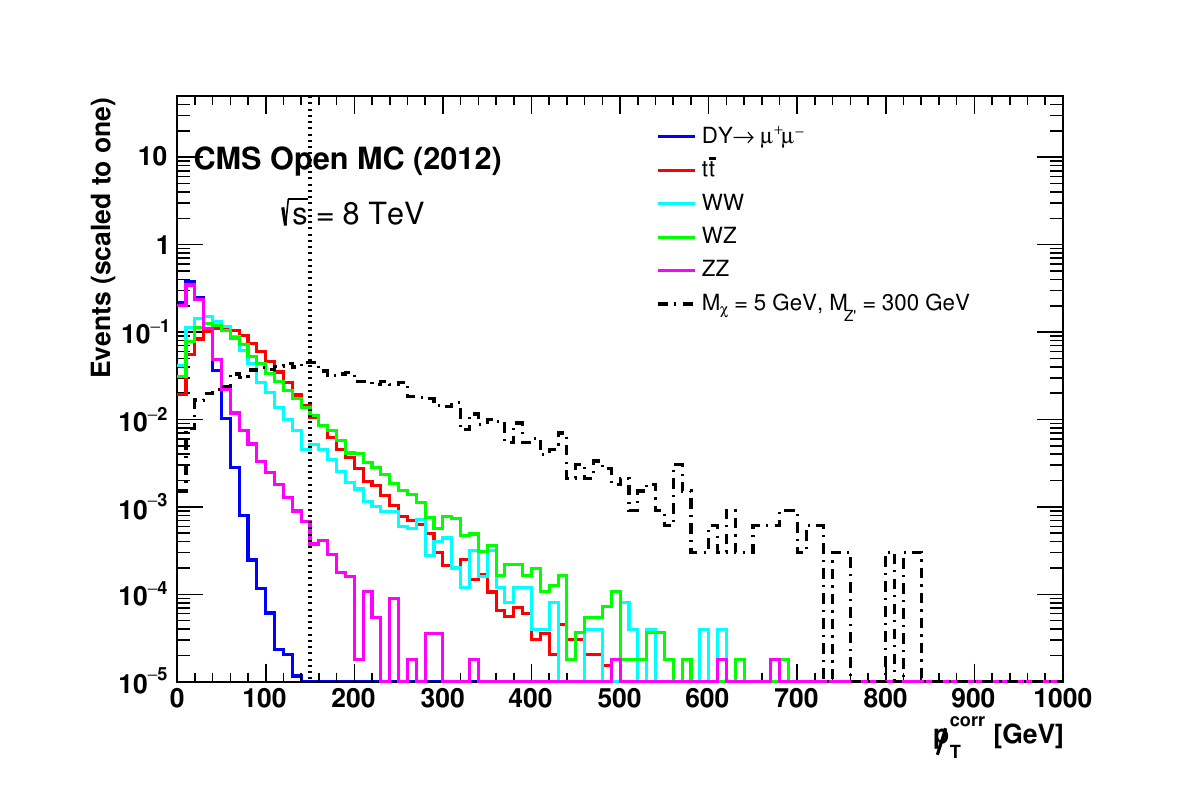}
  \label{ptmiss}
}
\subfigure[$|p_{T}^{\mu^{+}\mu^{-}} - \slashed{p}_{T}^{\text{corr}}|/p_{T}^{\mu^{+}\mu^{-}}$ ]{
  \includegraphics[width=85mm]{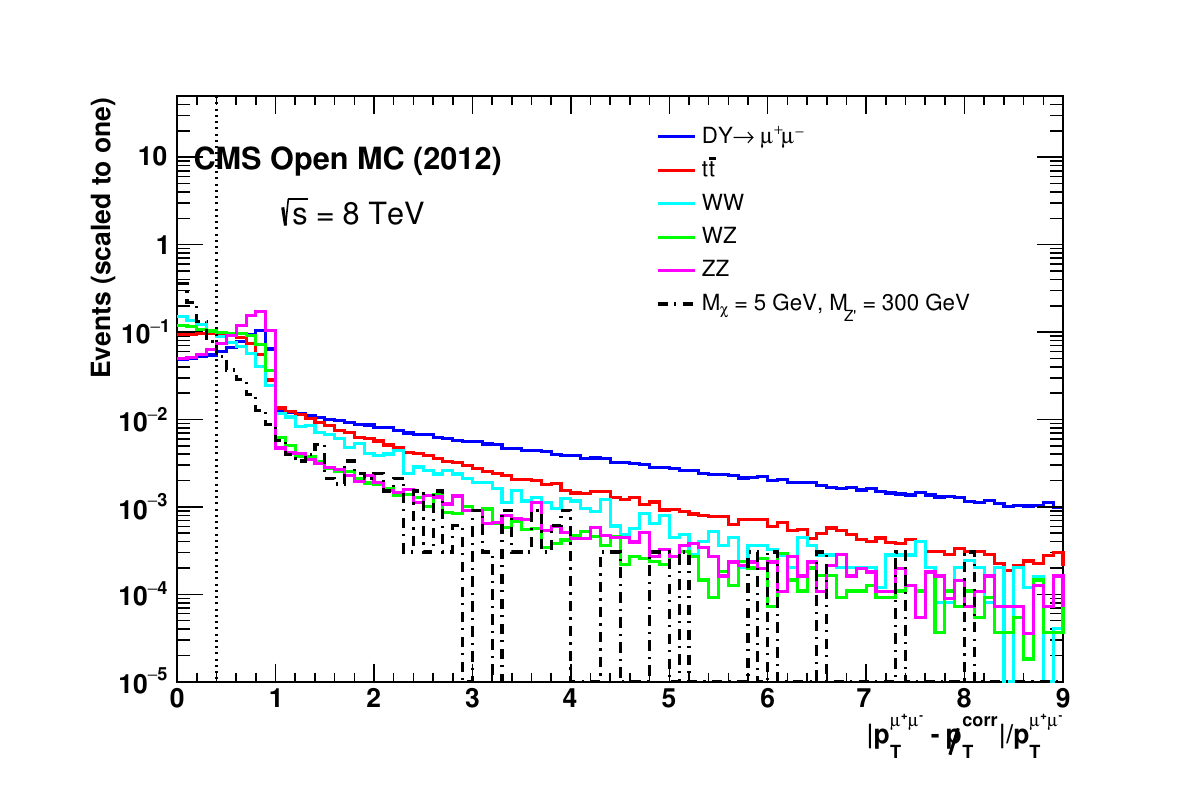}
  \label{ptdiff}
}
\hspace{0mm}
\subfigure[$\Delta\phi_{\mu^{+}\mu^{-},\vec{\slashed{p}}_{T}^{\text{corr}}}$]{
  \includegraphics[width=85mm]{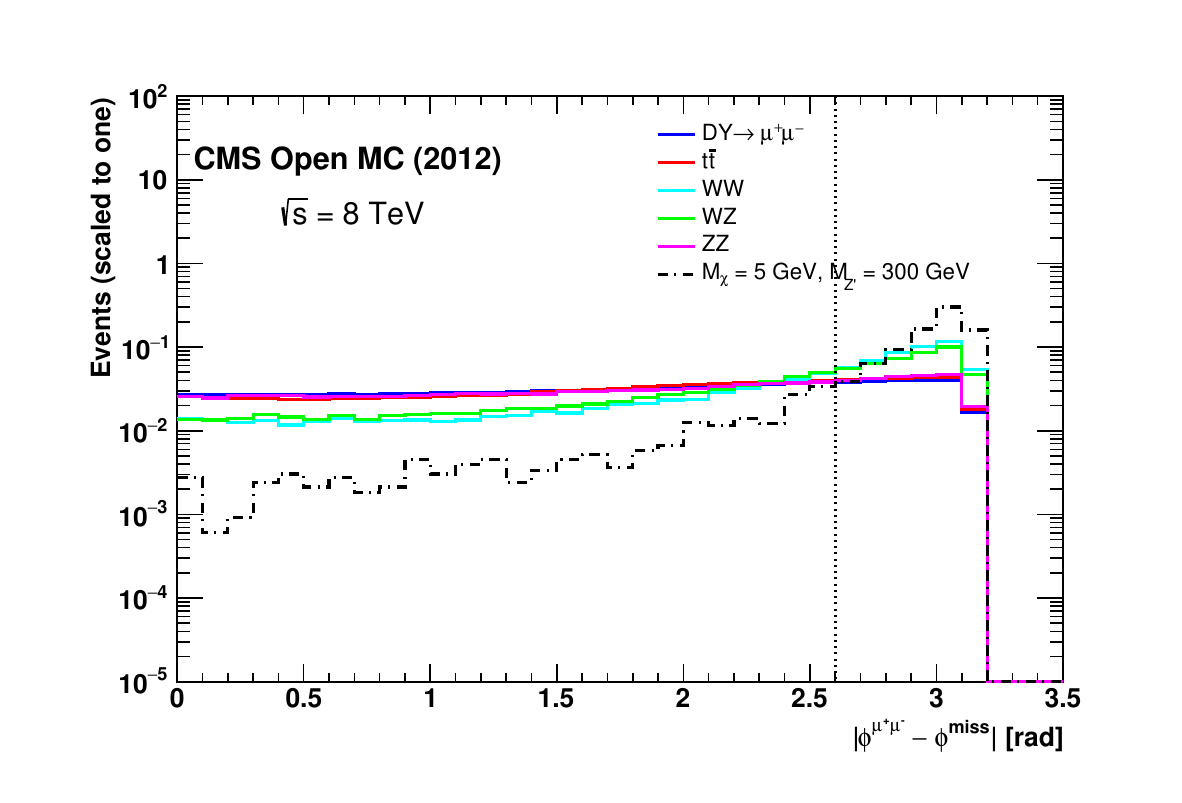}
  \label{deltaphi}
}
\caption{Distributions of $\slashed{p}_{T}^{\text{corr}}$ (\ref{ptmiss}), $|p_{T}^{\mu^{+}\mu^{-}} - \slashed{p}_{T}^{\text{corr}}|/p_{T}^{\mu^{+}\mu^{-}}$ (\ref{ptdiff}) and $\Delta\phi_{\mu^{+}\mu^{-},\vec{\slashed{p}}_{T}^{\text{corr}}}$ (\ref{deltaphi}) for the signal presentation of the model corresponding to the DH scenario with $M_{Z^{\prime}}=300$ GeV and SM backgrounds, for dimuon events with each muon passing the high $p_T$ muon ID discussed in section \ref{section:AnSelection}. The vertical dashed lines correspond to the chosen cut value per each variable. All histograms are normalized to unity to highlight qualitative features.}
\label{cuts}
\end{figure*}

In figure \ref{cuts}, for dimuon events where each muon passes the high $p_T$ muon ID discussed in section \ref{section:AnSelection}, we present the distributions of $\slashed{p}_{T}^{\text{corr}}$ (\ref{ptmiss}), $|p_{T}^{\mu^{+}\mu^{-}} - \slashed{p}_{T}^{\text{corr}}|/p_{T}^{\mu^{+}\mu^{-}}$ (\ref{ptdiff}) and $\Delta\phi_{\mu^{+}\mu^{-},\vec{\slashed{p}}_{T}^{\text{corr}}}$ (\ref{deltaphi}) for the signal presentation of the model corresponding to the DH scenario with $M_{Z^{\prime}}=300$ GeV, $M_{\chi} = 5$ GeV, $g_{SM} = 0.25$ and for SM backgrounds. 
The distributions are scaled to one. In these plots, the vertical dashed lines correspond to the chosen cut value for each variable.

Table \ref{table:selection2} summarizes the strict selection criteria. 
By using the tighter cuts in addition to those listed in selection 1 (summarized in table \ref{table:selection1}), we could fully suppress the DY and ZZ backgrounds, in addition to a significant reduction in the contributions of $t\bar{t}$, WW and WZ backgrounds.

\begin{table*}
    \centering
    \begin {tabular} {|c|c|c|}
\hline
step & variable & requirements \\

\hline
    & Trigger     & HLT\_Mu40\_eta2p1 \cite{HLT} \\
    & High $p_{T}$ muon ID & \cite{R41, R32}\\
Selection (1)    & $p^{\mu}_{T}$ (GeV) & $>$ 45 \\
    & $\eta^{\mu}$ (rad) & $<$ 2.1 \\
    &Mass window (GeV) & $(0.9 \times M_{Z^{\prime}}) < M_{\mu^{+}\mu^{-}} < (M_{Z^{\prime}} + 25)$ \\
    & $|\text{cos}\theta_{\text{CS}}|$& $\leq 0.8$ \\

    \hline
    &$\slashed{p}_{T}^{\text{corr}}$ (GeV) & $> 150$ \\   
Selection (2)     &$|p_{T}^{\mu^{+}\mu^{-}} - \slashed{p}_{T}^{\text{corr}}|/p_{T}^{\mu^{+}\mu^{-}}$ & $< 0.4$  \\
     &$\Delta\phi_{\mu^{+}\mu^{-},\vec{\slashed{p}}_{T}^{\text{corr}}}$ (rad) & $> 2.6$ \\
         \hline
    \end{tabular}
    \caption{Summary of final selection for the analysis.}
    \label{table:selection2}
\end{table*}

\begin{figure*}
\centering
\subfigure[180 $< M_{\mu^+\mu^-} <$ 225 GeV]{
  \includegraphics[width=60mm]{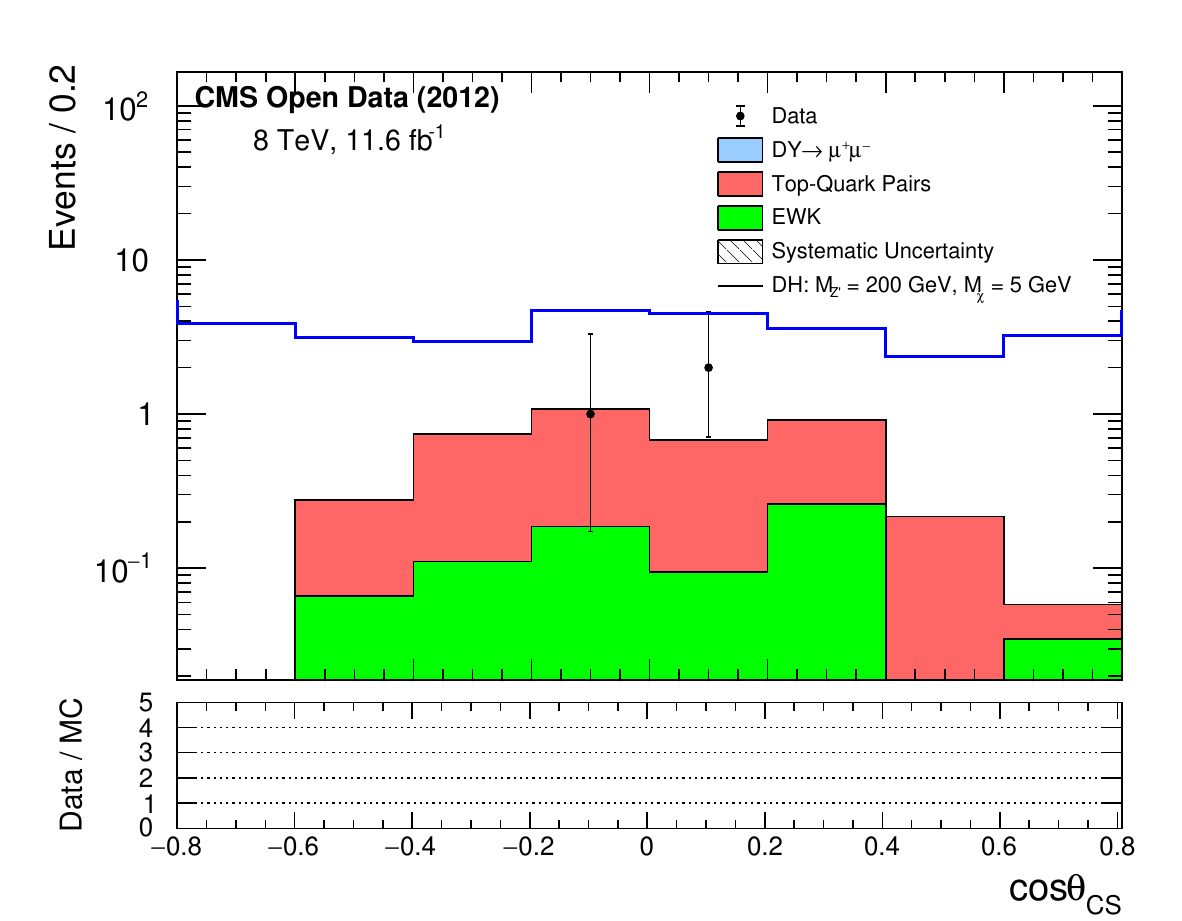}
  \label{bin200}
}
\subfigure[225 $< M_{\mu^+\mu^-} <$ 275 GeV]{
  \includegraphics[width=60mm]{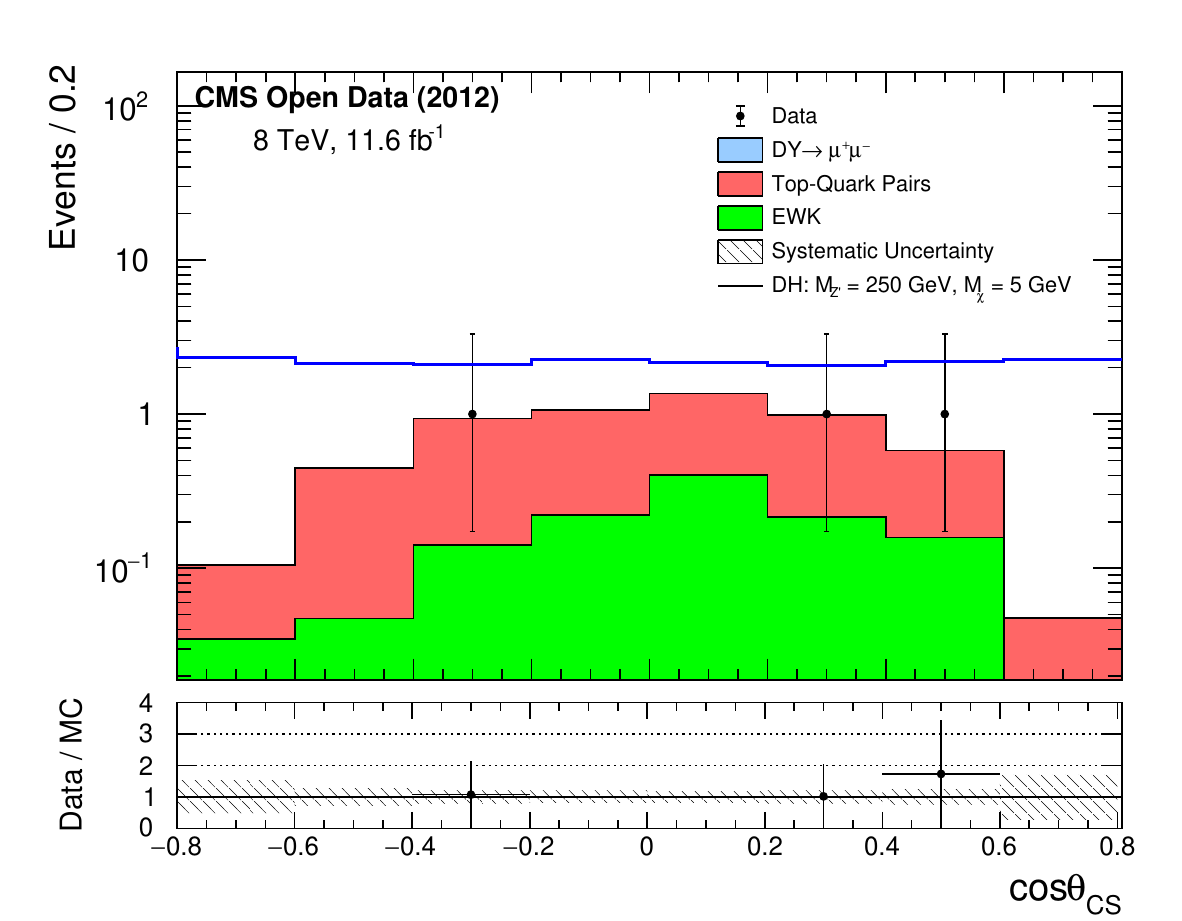}
  \label{bin250}
}
\hspace{0mm}
\subfigure[270 $< M_{\mu^+\mu^-} <$ 325 GeV]{
  \includegraphics[width=60mm]{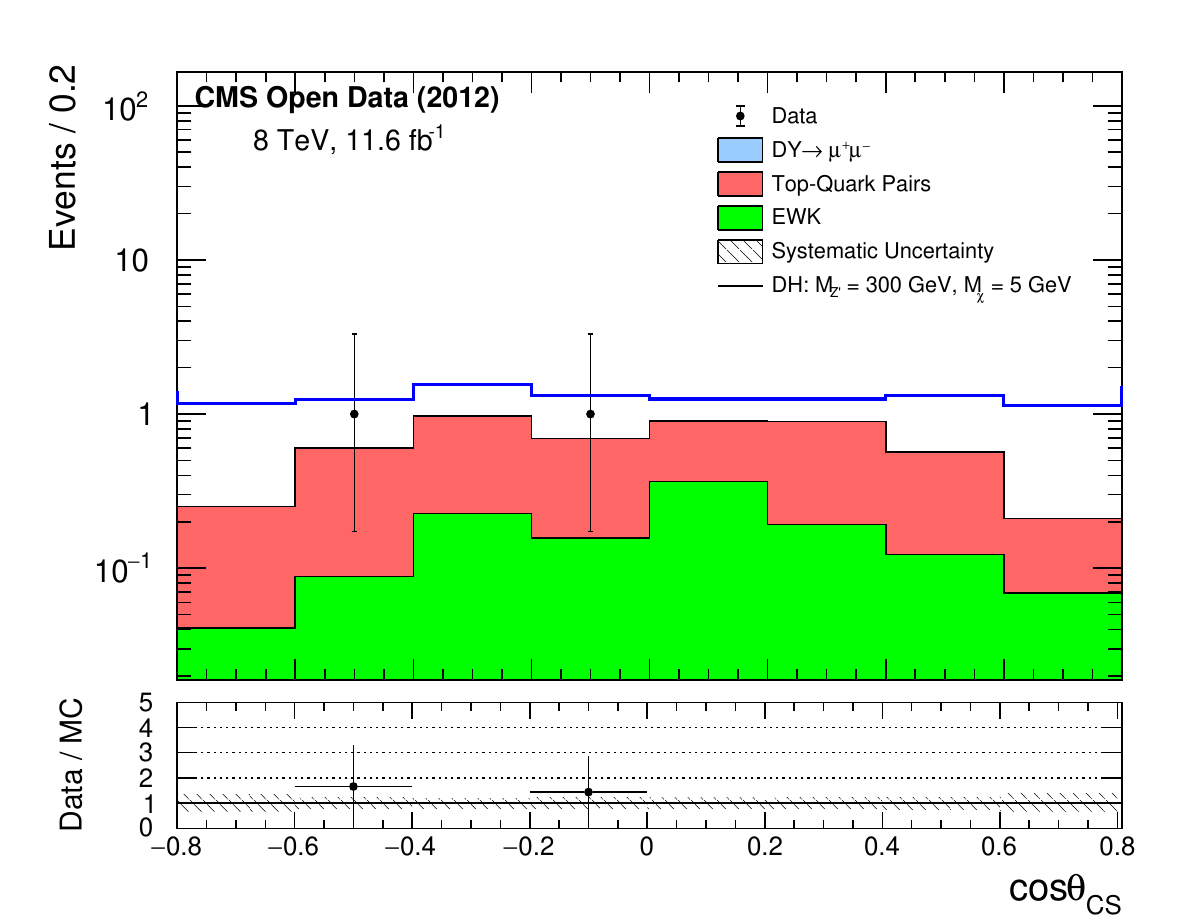}
  \label{bin300}
}
\subfigure[315 $< M_{\mu^+\mu^-} <$ 375 GeV]{
  \includegraphics[width=60mm]{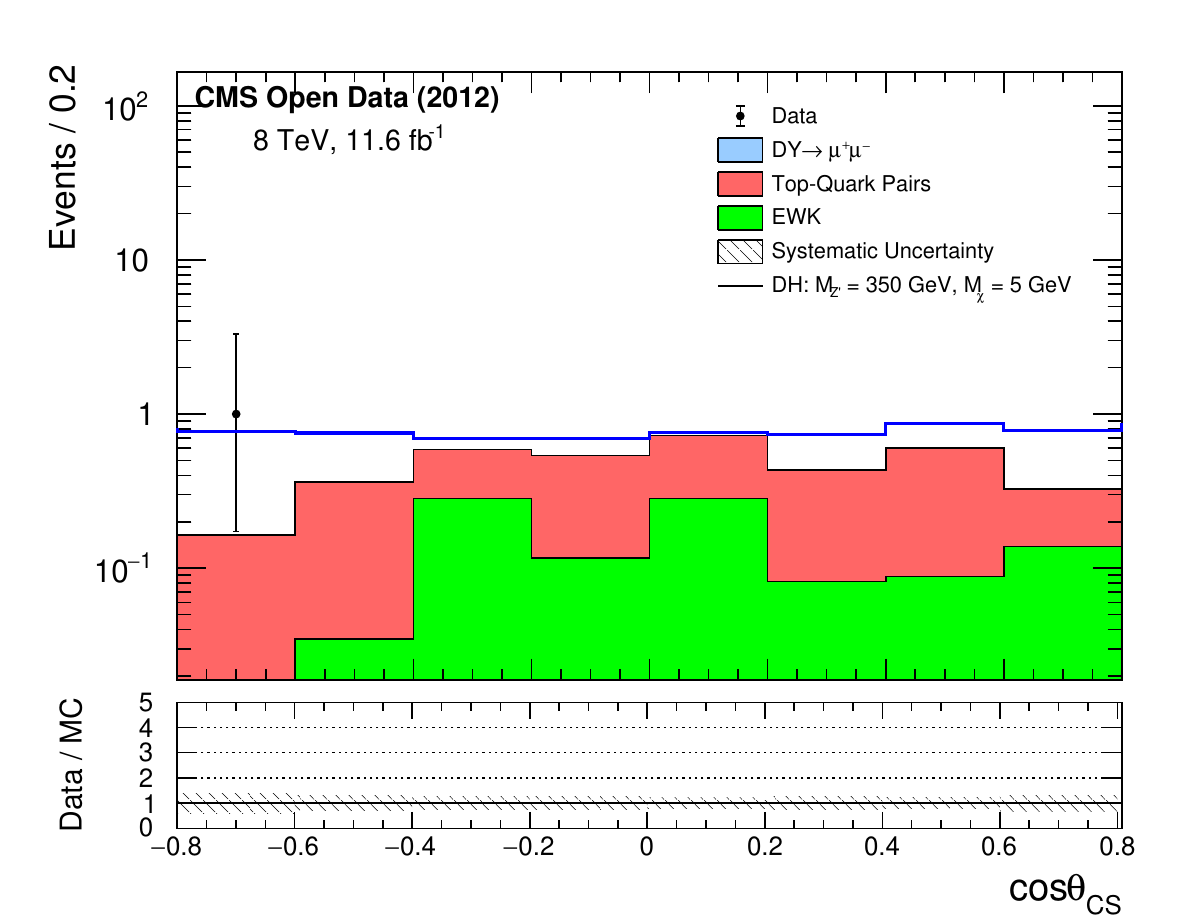}
  \label{bin350}
}
\hspace{0mm}
\subfigure[360 $< M_{\mu^+\mu^-} <$ 425 GeV]{
  \includegraphics[width=60mm]{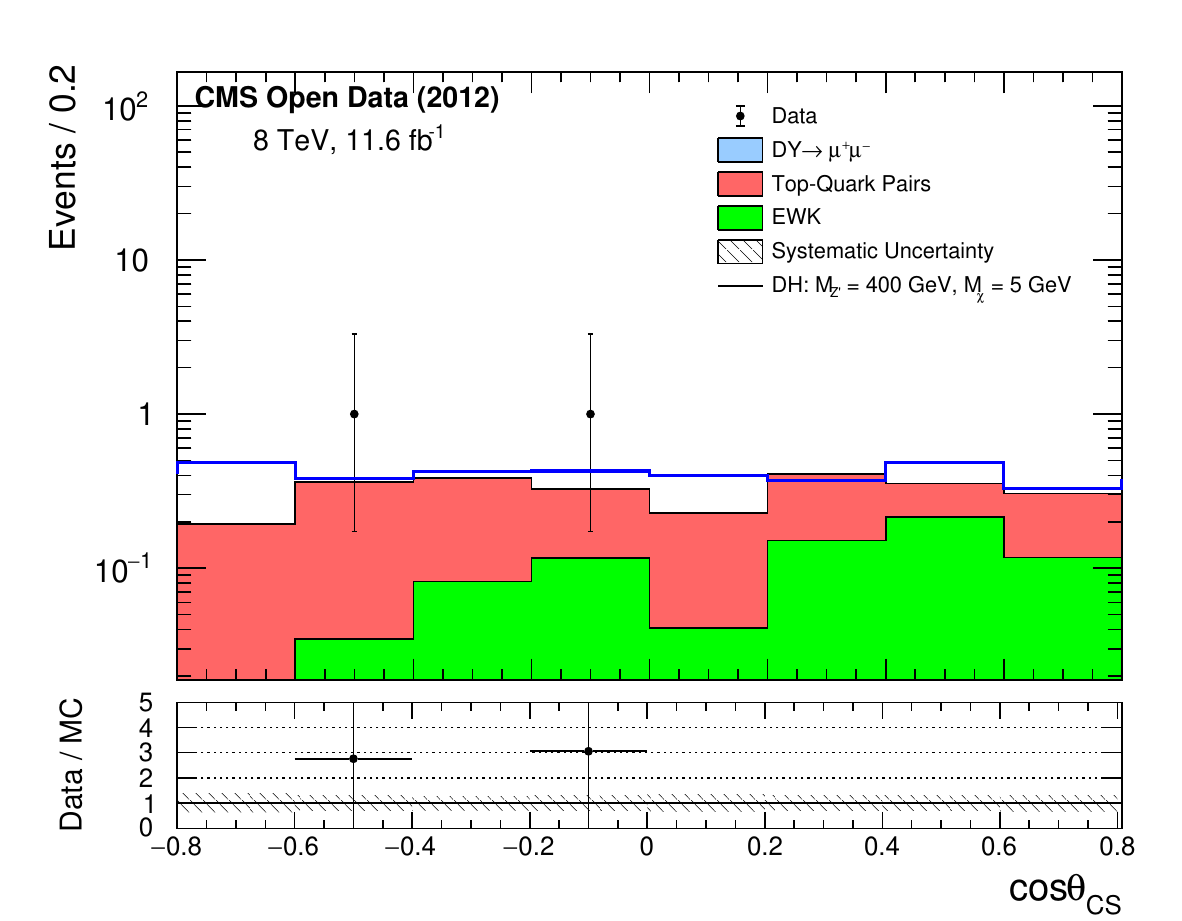}
  \label{bin400}
}
\subfigure[405 $< M_{\mu^+\mu^-} <$ 475 GeV]{
  \includegraphics[width=60mm]{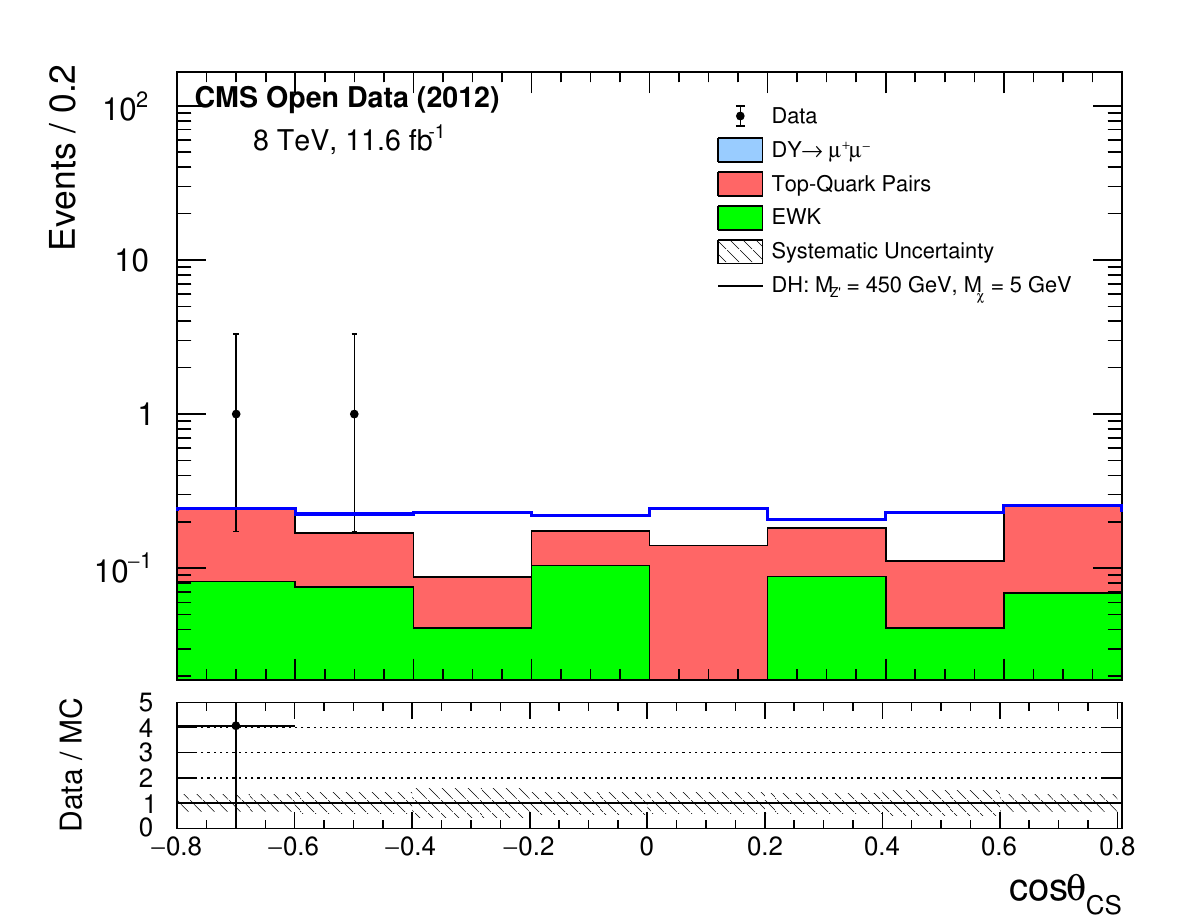}
  \label{bin450}
}
\hspace{0mm}
\subfigure[495 $< M_{\mu^+\mu^-} <$ 575 GeV]{
  \includegraphics[width=60mm]{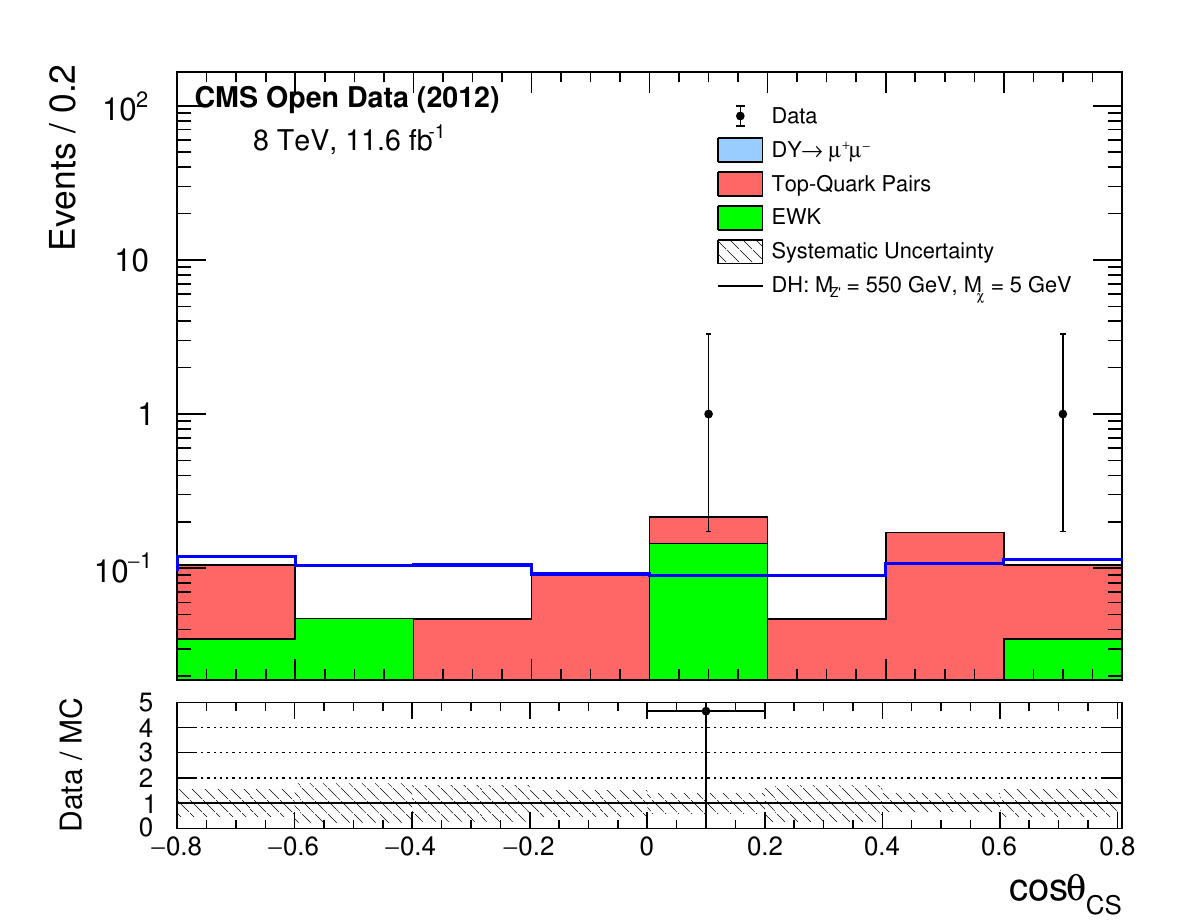}
  \label{bin550}
}
\subfigure[540 $< M_{\mu^+\mu^-} <$ 625 GeV]{
  \includegraphics[width=60mm]{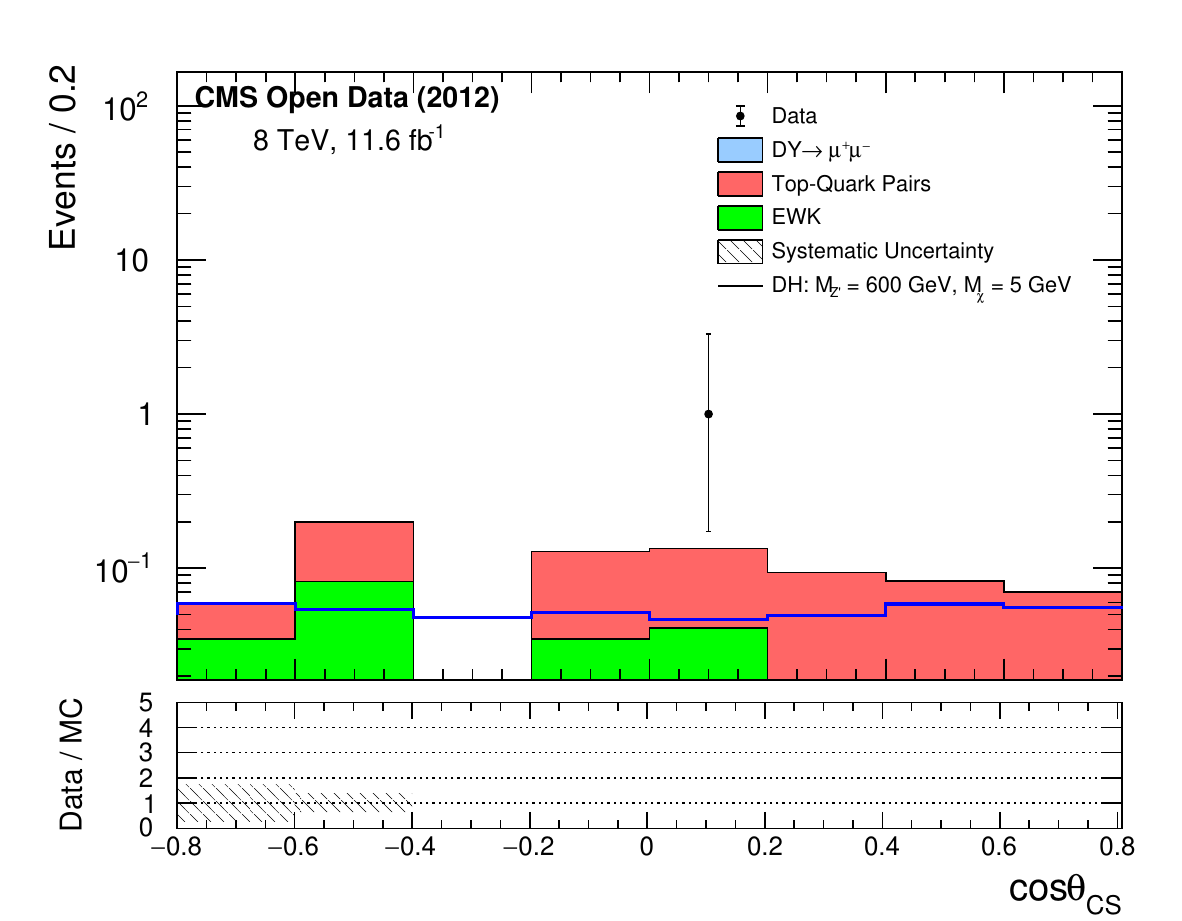}
  \label{bin600}
}
\hspace{0mm}
\caption{Distributions of cos$\theta_{CS}$ are illustrated, for events passing final selections listed in table \ref{table:selection2}, for data (dots) and Standard Model expectations (histograms) for several dimuon mass windows.
The signals presentation of the model corresponding to the DH scenario with the value of $M_{Z^{\prime}}$ runs from 200 to 600 GeV are superimposed.
In the lower bands, the data-to-simulation ratio is displayed. The shaded region represents the statistical and systematic uncertainties in the predicted backgrounds, added in quadrature.}
\label{fig8}
\end{figure*}

\begin{table*}
\centering
\tiny
\fontsize{6.5pt}{12pt}
\selectfont
\begin {tabular} {|l|c|c|c|c|c|c|c|c|c|c|c|}
\hline
$M_{Z^{\prime}}$ (GeV) & 200 & 250 & 300 & 350 & 400 & 450 & 500 & 550 & 600 & 700 \\
\hline
$\text{t}\bar{\text{t}} + \text{jets}$
& $3.21 \pm 0.47$ 
& $4.31 \pm 0.60$
& $3.84 \pm 0.55$
& $2.71 \pm 0.41$
& $1.80 \pm 0.30$
& $0.87 \pm 0.18$
& $0.44 \pm 0.12$
& $0.56 \pm 0.13$
& $0.56 \pm 0.13$
& $0.30 \pm 0.09$
\\
\hline
$\text{WW + jets}$ 
&$0.66 \pm 0.17$
&$1.11 \pm 0.24$
&$1.14 \pm 0.24$
&$0.97 \pm 0.22$
&$0.69 \pm 0.18$ 
&$0.45 \pm 0.14$
&$0.38 \pm 0.12$
&$0.24 \pm 0.10$
&$0.17 \pm 0.08$
&$0.21 \pm 0.09$
\\
\hline
$\text{WZ + jets}$ 
& $0.10 \pm 0.03$ 
& $0.11 \pm 0.03$
& $0.12 \pm 0.03$
& $0.06 \pm 0.02$
& $0.07 \pm 0.02$
& $0.05 \pm 0.02$
& $0.03 \pm 0.01$
& $0.03 \pm 0.01$
& $0.03 \pm 0.02$
& $0.03 \pm 0.01$
\\
\hline
Sum Bkgs 
&$3.97 \pm 0.56$ 
&$5.53 \pm 0.74$
&$5.10 \pm 0.70$ 
&$3.74 \pm 0.54$
&$2.56 \pm 0.40$
&$1.37 \pm 0.24$ 
&$0.85 \pm 0.18$
&$0.83 \pm 0.17$
&$0.76 \pm 0.16$
&$0.54 \pm 0.13$ 

\\
\hline
\hline
Data
&$3$ 
&$3$ 
&$2$
&$1$
&$2$
&$2$ 
&$0$ 
&$2$
&$1$
&$0$ 

\\
\hline
\hline
DH signal
&$28.36 \pm 3.60$
&$17.55 \pm 2.17$
&$10.26 \pm 1.26$
&$6.05 \pm  0.73$
&$3.31 \pm 0.41$
&$1.86 \pm 0.23$
&$1.15 \pm 0.14$ 
&$0.82 \pm 0.10$
&$0.42 \pm 0.05$
&$0.18 \pm 0.02$ 
\\
\hline
\end{tabular}
\caption{Table shows the number of events that meet the criteria of the final event selection listed in table \ref{table:selection2} for various $Z^{\prime}$ mass points within the DH scenario. Assumes the coupling constants of $g_{DM} = 1.0$ and $g_{SM} = 0.25$, and a fixed dark matter mass  
$(M_{\chi} = 5$ GeV). The yields include all relevant SM backgrounds and the CMS open data, corresponding to an integrated luminosity of 11.6 fb$^{-1}$. The total uncertainties, including both statistical and systematic components.}
\label{table:tab8}
\end{table*}
The graphs depicted in figure \ref{fig8} show the distributions of the cos$\theta_{CS}$ for events that have passed the event final selection criteria mentioned in tables \ref{table:selection2} in multiple mass bins:
180 $< M_{\mu^+\mu^-} <$ 225 GeV (\ref{bin200}), 
225 $< M_{\mu^+\mu^-} <$ 275 GeV (\ref{bin250}), 
270 $< M_{\mu^+\mu^-} <$ 325 GeV (\ref{bin300}), 
315 $< M_{\mu^+\mu^-} <$ 375 GeV (\ref{bin350}), 
360 $< M_{\mu^+\mu^-} <$ 425 GeV (\ref{bin400}), 
405 $< M_{\mu^+\mu^-} <$ 475 GeV (\ref{bin450}) 
495 $< M_{\mu^+\mu^-} <$ 575 GeV (\ref{bin550}) and 
540 $< M_{\mu^+\mu^-} <$ 625 GeV (\ref{bin600}). 
The shaded region in each graph indicates the statistical and systematic uncertainties, elaborated in section \ref{section:Uncertainties} and added in quadrature in the predicted backgrounds.

The information presented in table \ref{table:tab8} displays the number of events that meet the final event selection criteria for different $Z^{\prime}$ mass points based on the DH scenario. The coupling constants that are being assumed are $g_{DM} = 1.0$ and $g_{SM} = 0.25$, while the dark matter mass is fixed at $M_{\chi} = 5$ GeV. The yields that are presented in the table include all relevant SM backgrounds and the CMS open data, and the results correspond to an integrated luminosity of 11.6 fb$^{-1}$. 
The total uncertainties, including statistical and systematic components, are also considered in the table.

Applying the final cuts mentioned in table \ref{table:selection2} demonstrates that the signal samples are easily distinguishable from the SM backgrounds. 
Nevertheless, good agreement between data and the SM backgrounds is seen within the theoretical and systematic uncertainties for the studied mass bins, and no new physics has been observed.

\subsection{Statistical interpretation}
In order to interpret our results statistically, we conducted a statistical test using the profile likelihood method. We utilized the modified frequentist construction CLs \cite{R58, R59} in the asymptotic approximation \cite{R2} to establish exclusion limits on the product of signal cross sections and the branching fraction Br($Z^{\prime}$ $\rightarrow \mu\mu$) at a 95\% confidence level.

\begin{figure}
\centering
  \resizebox*{9.0cm}{!}{\includegraphics{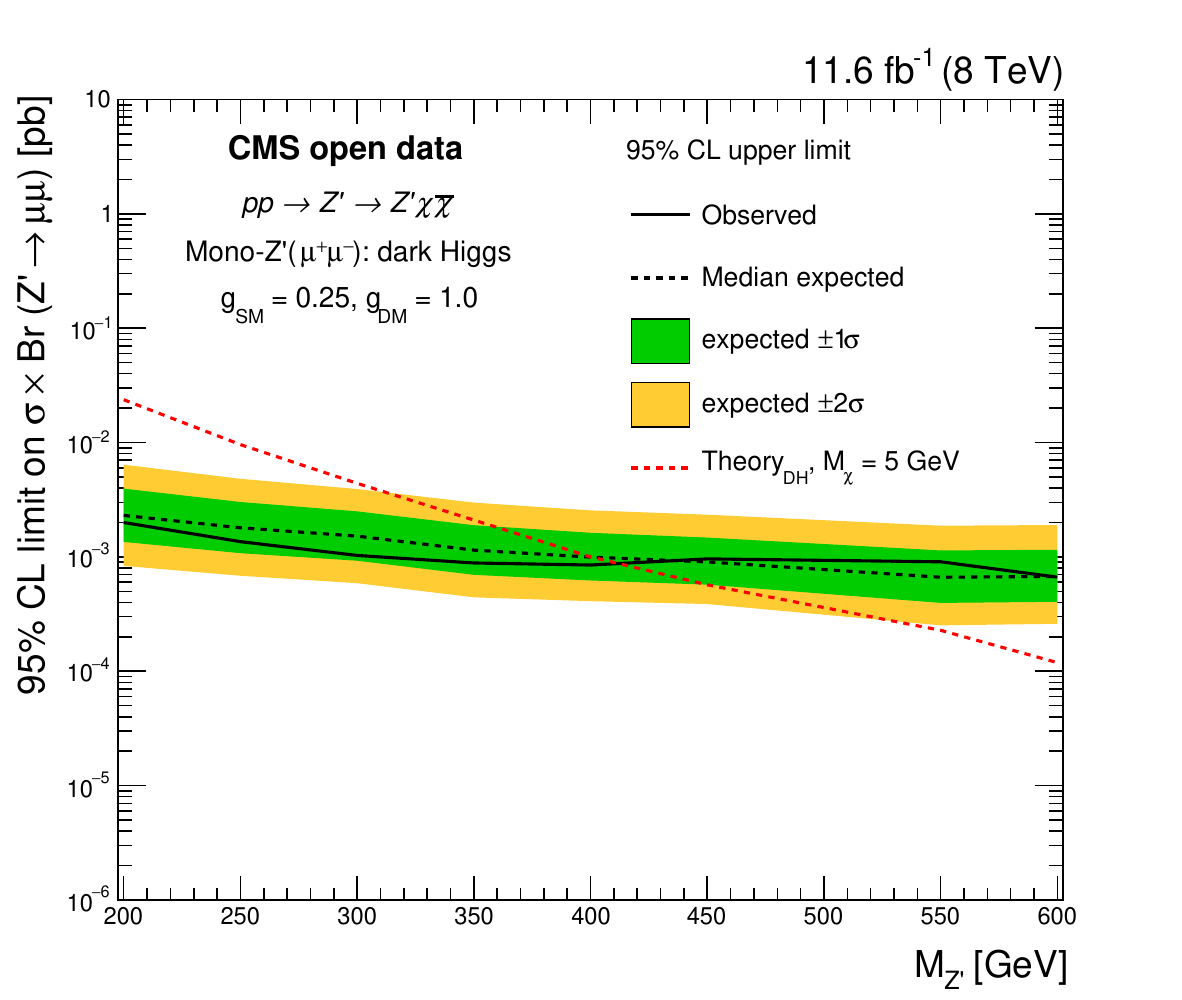}}
  \caption{95\% CL upper limits on the cross-section times the branching ratio (expected and observed), as a function of the mediator’s mass ($M_{Z^{\prime}})$, regarding the DH scenario, with the heavy dark sector set of masses and the muonic decay of the Z$^{\prime}$. The red dotted line represents the dark Higgs model at a fixed dark matter mass ($M_{\chi} = 5$ GeV).}
  \label{limit}
\end{figure}

In Figure \ref{limit}, the cross-section multiplied by the branching ratio Br($Z^{\prime}$ $\rightarrow \mu\mu$) sets a limit for the simplified model (DH). 
This includes the masses in the heavy dark sector, the muonic decay of the Z$^{\prime}$, and the coupling values of $\texttt{g}_{SM} = 0.25$ and $\texttt{g}_{DM} = 1.0$. The red dotted line on the graph represents the dark Higgs model at a specific dark matter mass ($M_{\chi} = $ 5 GeV).

In the DH scenario, we exclude $Z^{\prime}$ production with masses below 400 GeV for the expected median and 412 GeV for the observed data. 
The cross-section measurements multiplied by the $Z^{\prime}$ $\rightarrow \mu\mu$ branching ratios for the DH scenario, as listed in Table \ref{table:tabchi}, only vary with the change of both the mediator and dark Higgs masses and are independent of the choice of dark matter mass \cite{R1}. Consequently, the expected and observed limits on $M_{Z'}$ remain unchanged regardless of the DM mass alteration.

\section{Summary}
\label{section:Summary}
One approach to searching for new physics beyond the Standard Model at the LHC is to study changes in the dilepton mass spectrum at high mass. In addition, by analyzing the angular distributions of leptons, one can differentiate between different new physics models and measure the spin, parity, and couplings of any potential signal.

Our research examined the angular distributions of high-mass dimuon pairs in the Collins-Soper frame using datasets released by the CMS open data project. The data we used was obtained from proton-proton collisions during run 1, at $\sqrt{s} = 8$ TeV, corresponding to an integrated luminosity of 11.6 fb$^{-1}$. 
Our analysis focused on using the cos$\theta_{CS}$ variable to gain insights into the data.

We have compared the data with the expected outcomes of the Standard Model and found a good agreement for masses up to 600 GeV. In addition, we have compared the anticipated distributions of the cos$\theta_{CS}$ variable for high-mass Standard Model background events with that of a new physics scenario called the dark Higgs based on the Mono-Z$^{\prime}$ model. To improve distinguishing between signal events and SM backgrounds, we have applied powerful discrimination cuts that completely suppressed the DY and ZZ backgrounds. With these additional tight cuts, we observed good agreement between data and the SM backgrounds within the theoretical and systematic uncertainties, and no new physics has been discovered.

We have established a 95\% confidence level upper limit on certain model-free parameters. For the DH scenario with coupling values of $\texttt{g}_{SM} = 0.25$ and $\texttt{g}_{DM} = 1.0$, Z$^{\prime}$ boson masses below 400 GeV for the expected median and 412 GeV for the observed data have been ruled out.

\begin{acknowledgments}
The author thanks Tongyan Lin for her valuable contributions to the paper, including discussions on theoretical models, crosschecking results, and sharing Madgraph cards for event generation.
This paper is based on works supported by the Science, Technology, and Innovation Funding Authority (STDF) under grant number 48289.
\end{acknowledgments}

\nocite{*}

\end{document}